\documentclass[final,5p,times]{elsarticle}

\usepackage{amsfonts}
\usepackage{pifont}
\usepackage{appendix}
\usepackage{amsmath}
\usepackage{multirow}

\usepackage{graphics}
\usepackage{color}
\usepackage{algorithm}
\usepackage{algorithmicx}
\usepackage{algpseudocode}
\usepackage{mathrsfs}

\journal{Automatica}

\begin{document}

\begin{frontmatter}



\title{Power-Series Approach to Moment-Matching-Based Model Reduction of MIMO Polynomial Nonlinear Systems}


\author[First,Second]{Chao~Huang}
\author[Third,Forth]{Alessandro~Astolfi}

\address[First]{College of Electronic and Information Engineering, Tongji University, Shanghai, 200092, China}

\address[Second]{State Key Laboratory of Autonomous Intelligent Unmanned Systems, Shanghai, 200092, China}

\address[Third]{Department of Electrical and Electronic Engineering, Imperial College London, London SW7 2AZ, U.K.}

\address[Forth]{Dipartimento di Informatica, Sistemi e Produzione, Università di Roma “Tor Vergata,”
Rome 00133, Italy}

\cortext[corref]{Corresponding author. Email address: csehuangchao@tongji.edu.cn. This work was supported in part by the National Natural Science Foundation of China under Grant 62373282, Grant 62350003.}

\begin{abstract}
The model reduction problem for high-order multi-input, multi-output (MIMO) polynomial nonlinear systems based on moment matching is addressed. The technique of power-series decomposition is exploited: this decomposes the solution of the nonlinear PDE characterizing the center manifold into the solutions of a series of recursively defined Sylvester equations. This approach allows yielding nonlinear reduced-order models in very much the same way as in the linear case (e.g. analytically). Algorithms are proposed for obtaining the order and the parameters of the reduced-order models with precision of degree $\kappa$. The approach also provides new insights into the nonlinear moment matching problem: first, a lower bound for the order of the reduced-order model is obtained, which, in the MIMO case, can be strictly less than the number of matched moments; second, it is revealed that the lower bound is affected by the ratio of the number of the input and output channels; third, it is shown that under mild conditions, a nonlinear reduced-order model can always be constructed with either a linear state equation or a linear output equation. 
\end{abstract}

\begin{keyword}

Model reduction, Moment Matching, Polynomial Nonlinear Systems, Center Manifold, Sylvester Equation
\end{keyword}

\end{frontmatter}



\newdefinition{Def}{Definition}
\newdefinition{Asu}{Assumption}
\newtheorem{Thm}{Theorem}
\newtheorem{Cor}{Corollary}
\newdefinition{Rem}{Remark}
\newtheorem{Lem}{Lemma}
\newproof{Proof}{Proof}


\section{Introduction}

With an increasing need for improved performance and precision, as well as the advancement of modelling technology, mathematical models have become more and more complex. As a result, there has been a growing interest in model reduction: this involves creating a reduced-order (and therefore a simplified) mathematical model, while retaining as much of the original model's behavior as possible. Many model reduction approaches exist in the literature: for examples balanced truncation \cite{Moore1981Principal,Sorensen2002Sylvester}; Hankel-norm approximation \cite{GLOVER1984All,Fujimoto2005Nonlinear}; proper orthogonal decomposition \cite{Grepl2007Efficient,Willcox2002Balanced}; and moment matching (see the following paragraphs for a literature review). In this paper, our focus is on model reduction of polynomial nonlinear systems via moment matching.

Model reduction of linear systems by moment matching has a rich history, which can be traced back to the problem of Nevanlinna-Pick interpolation \cite{Antoulas1986On,ANTOULAS1990Solution,Byrnes2002A}. A well-known technique for linear moment matching is the Krylov projection (and its variants), see e.g., \cite{Grimme1995Model,Jaimoukha1997Implicitly,Antoulas2005Approximation,Gugercin2008Model}, which, after decades of development, has become a numerically stable method capable of guaranteeing asymptotic stability of the reduced-order model. Among the numerous results around the Kylov methods, \cite{Gallivan2004Sylvester} has noted that the Krylov projectors solve Sylvester equations, which turns out to be an important observation in the time-domain characterization of moment matching, as discussed in what follows.

Moment matching has traditionally been under taken in the frequency domain. In \cite{Astolfi2010Model}, a time-domain characterization of moment matching was proposed. It was revealed that the moment of a dynamical system, linear or nonlinear, can be linked to its steady-state response (if it exists) when the system is excited by an exogenous signal produced by an autonomous signal generator. The steady-state response is either described via the solution of a Sylvester equation in the linear case, or via the solution of a nonlinear partial differential equation (PDE) in the nonlinear case, see. e.g. \cite{Carr1981Applications,Isidori1990Output,Huang2004Nonlinear,Huang2024Identification}. The merits of the time-domain characterization are at least three-fold. First, it provides a convenient way to combine moment matching with the preservation of additional structural properties of the original model, such as passivity (in contrast to Krylov projection methods, which provide less intuitive model parametrizations). Second, it makes moment matching for general nonlinear systems tractable: this has led to the rapid development of nonlinear model reduction methods, including one-sided \cite{Scarciotti2016Model,SHAKIB2023Model,Moreschini2024Closed} and two-sided \cite{Ionescu2016Nonlinear,Simard2021Nonlinear,MORESCHINI2024Data} moment matching methods. Third, it has led to model reduction by data-driven moment matching \cite{SCARCIOTTI2017Data,SCARCIOTTI2020Data,MAO2024Data}, whereby one can estimate the moment online from experimental data and then perform the matching.

The Sylvester equation is a linear equation which is well-known to researchers and engineers: on its basis linear moment matching problems can be solved analytically. In contrast, the nonlinear moment matching problem relies on the solution of PDEs and it poses greater challenges. The distinguishing feature of the power-series approach, originally proposed in \cite{HUANG1990on,Huang1992An,Huang1994On} to address the nonlinear output regulation problem, lies in its ability to decompose the solution of the underlying PDE into the solutions of a series of recursively defined Sylvester equations (locally around the origin). This does not only allow deriving nonlinear reduced-order models in very much the same way as in the linear case (e.g. analytically), but also provides new insights into the center manifold and the nonlinear moment matching problem. The contributions of the proposed power-series approach in the framework of model reduction are as follows.

\begin{enumerate}
\item Two methods (Methods I and II) for moment matching with precision of degree \( \kappa \), as defined in Section II, are proposed, each providing an algorithm for computing the order and parameters of the reduced-order model.

\item Both Method I and Method II provide lower bounds for the order of the reduced-order model. An interesting finding is that the lower bound is affected by the ratio of the number of input and output channels. When the number of input channels does not exceed the number of output channels, a smaller lower bound is obtained by Method I; otherwise, Method II often provides a smaller lower bound.

\item The power-series approach provides flexibility in constructing the reduced-order models. In particular, under mild conditions, it is always possible to construct nonlinear reduced-order models which have either a linear state equation (plus a nonlinear output equation), or a linear output equation (plus a nonlinear state equation). 

\end{enumerate}

It is worth noting that the construction of nonlinear reduced-order models relies on the solution of the linear problem. A primary objective of model reduction is the determination of the minimal order of the reduced-order model. Two-sided moment matching is very successful in this respect: through the Loewner matrix framework, it has been shown that there generically exists a unique model of order $n$ that matches $n$ moments of the original model, see e.g., \cite{Gallivan2004Model,Ionescu2016Two}. In contrast, the characterization of the minimal order for the one-sided (usually the right side) moment matching problem, especially in the multi-input and multi-output (MIMO) settings, is not well-studied. Note that \cite{Gallivan2004Model,SHAKIB2023Time} have shown the existence, but non-uniqueness, of reduced-models of order $n$ to match $n$ moments for MIMO linear models. In this paper, we are able to explicitly construct models of order less than $n$ that matches $n$ $0$-moments of the original model for both MIMO linear systems and MIMO polynomial nonlinear systems. 

The rest of the paper is organized as follows. In Section \ref{sec:Preliminary} we review the time-domain characterization of the concept of moment and moment matching. Moreover, the notion of moment matching with precision of degree $\kappa$ is introduced. The power-series solution of the nonlinear PDE characterizing the center manifold is provided in Section \ref{sec:Preliminary}, too. In Section \ref{sec:Moment Matching of Linear Dynamics} we focus on the linear problem, i.e., on the linear approximation of the underlying system. Two methods are proposed, each providing a lower bound on the dimension of the reduced-order linear dynamics and an algorithm for computing its parameters. In Sections \ref{Moment Matching with the Precision of Degree kappa_infty} and \ref{Moment Matching with the Precision of Degree kappa} moment matching with precision of degree $\kappa$ is considered. Methods I and II are explored in Sections \ref{Moment Matching with the Precision of Degree kappa_infty} and \ref{Moment Matching with the Precision of Degree kappa}, respectively. Section \ref{sec:Numerical Examples} gives numerical validations of the obtained results, and Section \ref{sec:Conclusions and Future Works} concludes the paper.

\textbf{Notation:} The set of natural, real and complex numbers are denoted by $\mathbb N, \mathbb R$ and $\mathbb C$, respectively. $\mathbb C^{\rm 0}$ denotes the imaginary axis of the complex plane. The set $\{i,i+1,\cdots,j-1,j\}$, where $i,j\in\mathbb N$, is denoted by $\mathbb N_{i:j}$. Then, $\mathbb N_{i:\infty}$ is alternatively written as $\mathbb N_{i}$ for short. $\mathscr C^k$ and $\mathscr C^\infty$ denote the set of $k$-times continuously differentiable mappings and smooth mappings, respectively. We use $ A^\dag$ to denote the Moore-Penrose pseudo-inverse of $A$. ${\rm vec}\left(A\right)$ denotes the vectorization of $A$, i.e., ${\rm vec}\left(A\right) = {\left[ {A_1^{\rm T},A_2^{\rm T}, \cdots A_n^{\rm T}} \right]^{\rm T}}$, where $A_i,i=1,2,\cdots,n$ are the columns of $A$.

\section{Preliminary}\label{sec:Preliminary}

Consider the model to be reduced, governed by continuous-time nonlinear equations of the form
\begin{equation}\label{eq_preliminary_nldequ_x0}
\begin{array}{l}
\dot x^{\rm o}\left( t \right) = f^{\rm o}\left( {x^{\rm o}\left( t \right),u\left( t \right)} \right),\\
y^{\rm o}\left( t \right) = h^{\rm o}\left( {x^{\rm o}\left( t \right),u\left( t \right)} \right),
\end{array}
\end{equation}
where $x^{\rm o}\left(t\right)\in\mathbb R^{n^{\rm o}}$ is the state, $u\left(t\right)\in\mathbb R^m$ is the input, and $y^{\rm o}\left(t\right)\in\mathbb R^p$ is the output. $f^{\rm o}\left(\cdot,\cdot\right)$ and $h^{\rm o}\left(\cdot,\cdot\right)$ are $\mathscr C^\infty$ mappings. Consider also a reduced-order model of the form
\begin{equation}\label{eq_preliminary_nldequ_xi0}
\begin{array}{l}
\dot x \left( t \right) = f \left( {x \left( t \right),u\left( t \right)} \right),\\
y \left( t \right) = h \left( {x \left( t \right),u\left( t \right)} \right),
\end{array}
\end{equation}
where $x\left(t\right)\in\mathbb R^{n}$ is the state with $n<n^{\rm o}$, $u\left(t\right)\in\mathbb R^{m}$ is the input, $y\left(t\right)\in\mathbb R^{p}$ is the output, and $f\left(\cdot,\cdot\right)$, $h\left(\cdot,\cdot\right)$ are $\mathscr C^\infty$ mappings. Suppose, in addition, that $f^{\rm o}\left( {0,0} \right) = 0$, $h^{\rm o}\left( {0,0} \right) = 0$, $f\left( {0,0} \right) = 0$, $h\left( {0,0} \right) = 0$.

Consider finally a signal generator described by equations of the form
\begin{equation}\label{eq_preliminary_nldequ_v0}
\begin{array}{l}
\dot v\left( t \right) = s\left( {v\left( t \right)} \right),\\
u\left( t \right) = {\mathbf u}\left( {v\left( t \right)} \right),
\end{array}
\end{equation}
where $v\left(t\right)\in\mathbb R^{\sigma}$ is the state, and $s\left( \cdot \right)$ and ${\mathbf u}\left(\cdot\right)$ are $\mathscr C^\infty$ mappings. Suppose, in addition, that $s\left( 0\right)=0$ and ${\mathbf u}\left( 0\right)=0$. In what follows, we make two standing assumptions.

\begin{Asu}[Center Manifold]\label{asu_CM}
For the model (\ref{eq_preliminary_nldequ_x0}) with the signal generator (\ref{eq_preliminary_nldequ_v0}), there is a unique mapping ${\mathbf x}^{\rm o}\left(\cdot\right)$, locally defined in a neighborhood of $v=0$, which solves the PDE
\begin{equation}\label{invariant_manifold_x}
\frac{{\partial {\mathbf{x}^{\rm o}}\left( v \right)}}{{\partial v}}s\left( {v} \right)= f^{\rm o}\left( {{\mathbf{x}^{\rm o}}\left( v \right),{\mathbf{u}}\left( v \right)} \right).
\end{equation}
Similarly, for (\ref{eq_preliminary_nldequ_xi0}) and (\ref{eq_preliminary_nldequ_v0}), there is a unique mapping ${\mathbf x}\left(v\right)$, locally defined in a neighborhood of $v=0$, which solves the PDE
\begin{equation}\label{invariant_manifold_xi}
\frac{{\partial {{\mathbf{x}}}\left( v \right)}}{{\partial v}}s\left( v \right) = {f}\left( {{{\mathbf{x}}}\left( v \right),{\mathbf{u}}\left( v \right)} \right).
\end{equation}
\end{Asu}

Assumption \ref{asu_CM} implies that the interconnected system (\ref{eq_preliminary_nldequ_x0}) and (\ref{eq_preliminary_nldequ_v0}) possesses an invariant manifold (called the center manifold), described by the equation $x^{\rm o}={\mathbf x}^{\rm o}\left(v\right)$. In fact, for the center manifold to exist, one needs $\lambda \left( {\frac{{\partial {f^{\rm{o}}}}}{{\partial x}}\left( {0,0} \right)} \right) \cap \lambda \left( {\frac{{\partial s}}{{\partial v}}\left( 0 \right)} \right) = \emptyset$ and $\lambda \left( {\frac{{\partial s}}{{\partial v}}\left( 0 \right)} \right)\in\mathbb C^0$, where $\lambda\left(M\right)$ denotes the eigenvalues of the square matrix $M$. The same arguments apply to the interconnected system (\ref{eq_preliminary_nldequ_xi0}) and (\ref{eq_preliminary_nldequ_v0}).

\begin{Asu}[Observability]\label{asu_obs}
The signal generator (\ref{eq_preliminary_nldequ_v0}) is observable, i.e., for any pair of initial conditions $v^{\rm a}\left(0\right)$ and $v^{\rm b}\left(0\right)$, such that $v^{\rm a}\left(0\right)\neq v^{\rm a}\left(0\right)$, the corresponding output trajectories ${\mathbf u}\left(v^{\rm a}\left(\cdot\right)\right)$ and ${\mathbf u}\left(v^{\rm b}\left(\cdot\right)\right)$ are such that ${\mathbf u}\left(v^{\rm a}\left(\cdot\right)\right)-{\mathbf u}\left(v^{\rm b}\left(\cdot\right)\right)$ is not identically $0$.
\end{Asu}

\subsection{Moment Matching with Precision of Degree $\kappa$}

The notion of (nonlinear) moment in \cite{Astolfi2010Model} is defined as follows.

\begin{Def}[Moment]\label{def_moment}
Consider the model (\ref{eq_preliminary_nldequ_x0}) and the signal generator (\ref{eq_preliminary_nldequ_v0}). Under Assumptions \ref{asu_CM} and \ref{asu_obs}, the function
\begin{equation}\label{invariant_manifold_y}
\mathbf y^{\rm o}\left(v\right):= h^{\rm o}\left(\mathbf x^{\rm o}\left(v\right),\mathbf u\left(v\right)\right)
\end{equation}
with $\mathbf x^{\rm o}\left(\cdot\right)$ solution of Eq. (\ref{invariant_manifold_x}), is the moment of (\ref{eq_preliminary_nldequ_x0}) at $\left(\mathbf u\left(\cdot\right),s\left(\cdot\right)\right)$. The moment of (\ref{eq_preliminary_nldequ_xi0}) at $\left(\mathbf u\left(\cdot\right),s\left(\cdot\right)\right)$ is defined similarly by
\begin{equation}\label{invariant_manifold_yr}
\mathbf y\left(v\right):= h\left(\mathbf x\left(v\right),\mathbf u\left(v\right)\right).
\end{equation}
\end{Def}

Inspired by the notion of $\kappa$-th order approximate solution introduced in the study of the nonlinear output regulation problem \cite{Huang2004Nonlinear}, this paper considers the notion of moment matching with precision of degree $\kappa$, as follows.

\begin{Def}[Moment Matching with Precision of Degree $\kappa$]\label{def_multiplr_moment_match}
Under Assumptions \ref{asu_CM} and \ref{asu_obs}, the model (\ref{eq_preliminary_nldequ_xi0}) is said to be a reduced-order model at $\left(\mathbf u\left(\cdot\right),s\left(\cdot\right)\right)$ of the model (\ref{eq_preliminary_nldequ_x0}) with precision of degree $\kappa$, if
\begin{equation}\label{Eq_MMM}
\mathbf y\left( {{v}} \right) =\mathbf y^{\rm o} \left( {{v}} \right)+o\left( {{{\left\| v \right\|}^\kappa}} \right),
\end{equation}
In this case, (\ref{eq_preliminary_nldequ_xi0}) is said to match the moment of (\ref{eq_preliminary_nldequ_x0}) at
$\left(\mathbf u\left(\cdot\right),s\left(\cdot\right)\right)$ with precision of degree $\kappa$.
\end{Def}



\subsection{Power-Series Solution to the PDEs}

Before introducing the power-series solution to the above PDEs, we introduce some notation. Given a vector $v\in\mathbb R^n$, its successive Kronecker product is defined as ${v^{\left( 0 \right)}} = 1, {v^{\left( 1 \right)}} = v$, and
\[{v^{\left( i \right)}} = \underbrace {v \otimes v \otimes  \cdots  \otimes v}_{i\quad \rm factors},\quad i\in\mathbb N_2.\]
Then $v^{\left[i\right]}$ is obtained from $v^{\left(i\right)}$ by removing the repeating elements in $v^{\left(i\right)}$ and arranging the remaining elements in the original order (since $v$ is arbitrary, such an operation is independent of $v$, see \cite{Huang2004Nonlinear} for an example). Note that $v^{\left(i\right)}$ and $v^{\left[i\right]}$ are of dimension $n^i$ and $\mathcal C_{n-1+i}^i$, respectively \cite{Huang2004Nonlinear}, where ${\cal C}_{n - 1 + i}^i = \frac{{\left( {n - 1 + i} \right)!}}{{i!\left( {n - 1} \right)!}}$, that is $n-1+i$ choose $i$. Moreover, for each $i$ and $n$, there exist matrices $M_i^n\in\mathbb R^{\mathcal C_{n-1+i}^i\times n^i}$ and $N_i^n\in\mathbb R^{n^i\times \mathcal C_{n-1+i}^i}$ independent of $v$, such that
\begin{equation}\label{MN}
{v^{\left[ i \right]}} = {M_i^n}{v^{\left( i \right)}},\quad {v^{\left( i \right)}} = {N_i^n}{v^{\left[ i \right]}}.
\end{equation}
Given a matrix $A\in\mathbb R^{n\times m}$, the successive Kronecker product $A^{\left(i\right)}$ is defined in the same way as $v^{\left(i\right)}$, and $A^{\left[i\right]}$ is defined as ${A^{\left[ i \right]}} = {M_i^n}{A^{\left( i \right)}}$. 

Given two square matrices $A\in\mathbb R^{n\times n}$ and $B\in\mathbb R^{m\times m}$, the Kronecker sum of $A$ and $B$ is defined as $A \oplus B = {I_m} \otimes A + B \otimes {I_n}$ \cite{Bellman1997Introduction}. The successive Kronecker sum is defined as ${A^{\left\{ 0 \right\}}} = I_n,\quad {A^{\left\{ 1 \right\}}} = A$, and
\[{A^{\left\{ i \right\}}} = \underbrace {A \oplus A \oplus  \cdots  \oplus A}_{i\quad \rm factors},\quad i \in\mathbb N_2. \]
It is shown in e.g. \cite{Huang2024Identification} that for a square matrix $A$, the above definition is equivalent to ${A^{\left\{ i \right\}}} = \sum\nolimits_{r = 1}^i I_{n}^{(r-1)} \otimes A \otimes I_{n}^{(i-r)}$ for $i\ge 1$, then for a non-square matrix $A$, we define ${A^{\left\{ i \right\}}}$ by this equation. 

\begin{Lem}[\cite{Huang2004Nonlinear}] \label{lemma_tmatrix}
Suppose that $A\in\mathbb R^{n\times n}$ is diagonalizable, and the spectrum of $A$ is ordered as $\lambda_1,\cdots,\lambda_{n}$. Then ${A^{\left\langle i \right\rangle }}=M_i^n{A^{\left\{ i \right\}}}N_i^n, i\in\mathbb N_1$ is diagonalizable, and its spectrum is given by
\[{\Lambda _i} = \left\{ \left. \lambda  \right|\lambda  = \sum\nolimits_{r = 1}^i {{\lambda_{{\rho _r}}}} ,{\rho _r} \in\mathbb N_{1:n} \right\}.\]
\end{Lem}

Suppose that the reduced-order model (\ref{eq_preliminary_nldequ_xi0}) is in polynomial form, i.e.,
\begin{equation}\label{eq_f}
f\left( {x,u} \right) = \sum\nolimits_{l = 1}^L {\sum\nolimits_{\scriptstyle i + r = l,\hfill\atop
\scriptstyle i,r \in\mathbb N\hfill} {{{\mathbf{F}}_{i,r}}} \left( {{x^{\left[ i \right]}} \otimes {u^{\left[ r \right]}}} \right)} 
\end{equation}
and
\begin{equation}\label{eq_h}
h\left( {x,u} \right) = \sum\nolimits_{l = 1}^L {\sum\nolimits_{\scriptstyle i + r = l,\hfill\atop
\scriptstyle i,r \in\mathbb N\hfill} {{{\mathbf{H}}_{i,r}}} \left( {{x^{\left[ i \right]}} \otimes {u^{\left[ r \right]}}} \right)} ,
\end{equation}
where $L\in\mathbb N_1$ ($L=\infty$ included provided that the series on the right-hand side of the equations converge for all $\left(x,u\right)$) is the maximum degree of nonlinearity, and ${\mathbf{F}}_{i,r}$ and ${\mathbf{H}}_{i,r}$ are matrices of appropriate dimensions. Following the notational convention of linear-systems, we denote $A = {\mathbf F}_{1,0},B = {\mathbf F}_{0,1},C =  {\mathbf H}_{1,0},D =  {\mathbf H}_{0,1}$.


\begin{Rem}
All notation for the model (\ref{eq_preliminary_nldequ_x0}) and the reduced-order model (\ref{eq_preliminary_nldequ_xi0}) follows this convention: any parameter of the original model is denoted with a superscript ``${\rm o}$", while the reduced-order model (\ref{eq_preliminary_nldequ_xi0}) does not use superscript. The meanings of the notation are otherwise the same for both (\ref{eq_preliminary_nldequ_x0}) and (\ref{eq_preliminary_nldequ_xi0}). For example, $\left(A^{\rm o}, B^{\rm o}, C^{\rm o}, D^{\rm o}\right)$ represents the linear part of the model (\ref{eq_preliminary_nldequ_x0}), while $\left(A, B, C, D\right)$ represents the linear part of the reduced-order model (\ref{eq_preliminary_nldequ_xi0}).
\end{Rem}

The signal generator (\ref{eq_preliminary_nldequ_v0}) is also assumed to be in polynomial form, i.e.,
\begin{equation}\label{power_series3}
\begin{array}{l}
s\left( v \right) = \sum\nolimits_{l = 1}^L {{S_l}{v^{\left[ l \right]}}} ,\\
{\bf{u}}\left( v \right) = \sum\nolimits_{l = 1}^L {{{\bf{U}}_l}{v^{\left[ l \right]}}} 
\end{array}
\end{equation}
where $S_l\in\mathbb R^{\sigma\times {\mathcal C}_{\sigma  - 1 + l}^l}$ and ${\mathbf{U}}_l\in\mathbb R^{m\times {\mathcal C}_{\sigma  - 1 + l}^l}, l=1,2,\cdots,L$. $S_l$ and ${\mathbf{U}}_l$ are assumed known. The following assumption is satisfied for the signal generator (\ref{eq_preliminary_nldequ_v0}).


\begin{Asu}\label{asu_spectrumS1}
The eigenvalues of $S_1$ are all semi-simple and located on the imaginary axis.
\end{Asu}

\begin{Rem}
Assumption \ref{asu_spectrumS1} implies that $S_1 $ may have repeated eigenvalues but is nonetheless diagonalizable in the complex domain. This, in fact, makes both \( \mathbf{y}^{\rm o}(\cdot) \) and \( \mathbf{y}(\cdot) \) $0$-moments. While $0$-moment matching may appear restrictive for linear model reduction, it is natural as far as nonlinear model reduction is concerned, for detail see \cite{Astolfi2010Model}.
\end{Rem}

We have the following result on the power-series solution to the PDE (\ref{invariant_manifold_xi}) and (\ref{invariant_manifold_yr}). This problem was originally studied in \cite{Huang2004Nonlinear} in the framework of nonlinear output regulation, and then \cite{Huang2024Identification} made some refinement in the framework of nonlinear system identification problem.

\begin{Lem}[\cite{Huang2024Identification}]\label{lem_stability}
The infinite power series
\begin{equation}\label{power_series1}
{\mathbf{x}}\left( v \right) = \sum\nolimits_{l = 1}^\infty {{{\mathbf{X}}_l}{v^{\left[l\right]}}} ,
\end{equation}
\begin{equation}\label{power_series2}
{\mathbf{y}}\left( v \right) = \sum\nolimits_{l = 1}^\infty {{{\mathbf{Y}}_l}{v^{\left[l\right]}}} ,
\end{equation}
formally satisfy Eqs. (\ref{invariant_manifold_xi}) and (\ref{invariant_manifold_yr}), respectively, if the pair $\left(\mathbf X_l,\mathbf Y_l\right)$ is the solution of the recursive Sylvester equations:

\begin{enumerate}

\item for $l=1$,
\begin{equation}\label{Sylvester_l1}
{{\mathbf{X}}_1}S_1 = A{{\mathbf{X}}_1} + B{{\mathbf{U}}_1},
\end{equation}
\begin{equation}\label{Sylvester_l2}
{{\mathbf{Y}}_1} = C{{\mathbf{X}}_1} + D{{\mathbf{U}}_1},
\end{equation}

\item for $l\in\mathbb N_2$,
\begin{equation}\label{Sylvester_l3}
{{\mathbf{X}}_l}{S_1^{\left\langle l \right\rangle }}= A{{\mathbf{X}}_l} + B{{\mathbf{U}}_l} + {{\mathbf{E}}_l} + {{\mathbf{F}}_l}{{\mathbf{W}}_l}  ,
\end{equation}
\begin{equation}\label{Sylvester_l4}
{{\mathbf{Y}}_l} = C{{\mathbf{X}}_l} + D{{\mathbf{U}}_l}  + {{\mathbf{G}}_l} + {{\mathbf{H}}_l}{{\mathbf{W}}_l} ,
\end{equation}
where
\begin{equation}\label{Wl}
\begin{array}{l}
{{\mathbf{F}}_l} = \left[ {\begin{array}{*{20}{c}}
{{{\mathbf{F}}_{l,0}}}&{{{\mathbf{F}}_{l - 1,1}}}& \cdots &{{{\mathbf{F}}_{0,l}}}
\end{array}} \right],\\
{{\mathbf{H}}_l} = \left[ {\begin{array}{*{20}{c}}
{{{\mathbf{H}}_{l,0}}}&{{{\mathbf{H}}_{l - 1,1}}}& \cdots &{{{\mathbf{H}}_{0,l}}}
\end{array}} \right],\\
{{\mathbf{W}}_l} = {\left[ {\begin{array}{*{20}{c}}
{{\mathbf{W}}_{l,0}^{\rm{T}}}&{{\mathbf{W}}_{l - 1,1}^{\rm{T}}}& \cdots &{{\mathbf{W}}_{0,l}^{\rm{T}}}
\end{array}} \right]^{\rm{T}}},
\end{array}
\end{equation}
with
\begin{equation}\label{Wir}
{{\mathbf{W}}_{i,r}} = \left( {{\mathbf{X}}_1^{\left[ i \right]} \otimes {\mathbf{U}}_1^{\left[ r \right]}} \right)N_l^\sigma , \quad i+r=l, \quad i,r\in\mathbb N,
\end{equation}
and
\begin{equation}\label{El}
{{\mathbf{E}}_l} = \left\{ {\begin{array}{*{20}{l}}
{ - {{\mathbf{X}}_1}{S_2},\quad l = 2,}\\
\begin{array}{l}
\left[ {\sum\limits_{s = 2}^{l - 1} {\left( {\sum\limits_{\scriptstyle i + r = s,\hfill\atop
\scriptstyle i,r \in \mathbb N\hfill} {{{\mathbf{F}}_{i,r}}\left( {M_i^n \otimes M_r^m} \right)\Delta _{l - s}^{i,r}} } \right)} } \right.\\
\left. { - {{\mathbf{X}}_s}M_s^\sigma {{\left( {{S_{l - s + 1}}M_{l - s + 1}^\sigma } \right)}^{\left\{ s \right\}}}} \right]N_l^\sigma  - {{\mathbf{X}}_1}{S_l},
\end{array}\\
{l \in\mathbb N_3, }
\end{array}} \right.
\end{equation}
\begin{equation}\label{Gl}
{{\mathbf{G}}_l} = \left\{ {\begin{array}{*{20}{l}}
{0,\quad l = 2,}\\
{\left[ {\sum\limits_{s = 2}^{l - 1} {\sum\limits_{\scriptstyle i + r = s,\hfill\atop
\scriptstyle i,r \in\mathbb N\hfill} {{{\mathbf{H}}_{i,r}}\left( {M_i^{n} \otimes M_r^{m}} \right){\Delta ^{i,r}_{l - s}}} } } \right]{N_l^{\sigma}}, }\\
l \in\mathbb N_3,
\end{array}} \right.
\end{equation}
with
\[
\Delta _\rho ^{i,r} = \left\{ {\begin{array}{*{20}{l}}
{{\delta _{i,\rho }},\quad r = 0,\quad i > 0,}\\
{{\gamma _{r,\rho }},\quad i = 0,\quad r > 0,}\\
{\sum\nolimits_{k = 0}^\rho  {{\delta _{i,k}} \otimes {\gamma _{r,\rho  - k}}} ,\quad i > 0,\quad r > 0,}
\end{array}} \right.
\]
\[\begin{array}{l}
{\delta _{i,\rho }} = \sum\limits_{\scriptstyle{\rho _1} +  \cdots  + {\rho _i} = i + \rho ,\hfill\atop
\scriptstyle{\rho _1}, \cdots ,{\rho _i} \ge 1\hfill} {\left( {{{\mathbf{X}}_{{\rho _1}}}M_{{\rho _1}}^\sigma } \right) \otimes  \cdots  \otimes \left( {{{\mathbf{X}}_{{\rho _i}}}M_{{\rho _i}}^\sigma } \right)} ,\\
i \ge 1,\quad \rho  \ge 0,
\end{array}\]
\[\begin{array}{l}
{\gamma _{r,\rho }} = \sum\limits_{\scriptstyle{\rho _1} +  \cdots  + {\rho _r} = r + \rho ,\hfill\atop
\scriptstyle{\rho _1}, \cdots ,{\rho _r} \ge 1\hfill} {\left( {{{\mathbf{U}}_{{\rho _1}}}M_{{\rho _1}}^\sigma } \right) \otimes  \cdots  \otimes \left( {{{\mathbf{U}}_{{\rho _r}}}M_{{\rho _r}}^\sigma } \right)} ,\\
r \ge 1,\quad \rho  \ge 0.
\end{array}\]
\end{enumerate}
Moreover, if Assumption \ref{asu_spectrumS1} holds and $A$ is Hurwitz, the solution $\left(\mathbf X_l,\mathbf Y_l\right)$ is unique for every $l\in\mathbb N_1$.
\end{Lem}

As stated in \cite{Huang2004Nonlinear}, if (\ref{power_series1}) converges in the neighborhood of $v=0$, then it is the $\mathscr C^\infty$ solution of Eq. (\ref{invariant_manifold_xi}). Moreover, due to the $\mathscr C^\infty$ nature of $\dot x =f\left(x,\mathbf u\left(v\right)\right)$ and $\dot v=s\left(v\right)$, according to \cite{Shilnikov2001Methods}, for any $k\in\mathbb N_1$, Eq. (\ref{invariant_manifold_xi}) has a $\mathscr C^k$ solution in a neighborhood of $v=0$ (the neighborhood may become smaller as $k$ increases), which has the same Taylor expansion at $v=0$ with that of (\ref{power_series1}). Similar arguments can be extended to the output equations (\ref{invariant_manifold_yr}) and (\ref{power_series2}) as well. For simplicity of notation, we impose the following assumption.

\begin{Asu}\label{asu_power_series_converge}
The power series (\ref{power_series1}) and (\ref{power_series2}) converge in a neighborhood of $v=0$.
\end{Asu}

According to \cite{Huang2024Identification}, suppose $P_l\in\mathbb C^{{\mathcal C}_{\sigma  - 1 + l}^l\times {\mathcal C}_{\sigma  - 1 + l}^l}$, where $l\in\mathbb N_1$, is such that
\[P_l^{ - 1}{S_1^{\left\langle l \right\rangle }}{P_l} = {\rm diag}\left( {{\lambda _{l,1}}, \cdots ,{\lambda _{l,{\mathcal C}_{\sigma  - 1 + l}^l}}} \right),\]
where $\lambda_{l,\ell}$ is the $\ell$-th eigenvalue of $S_1^{\left\langle l \right\rangle }$, and
\[{\Sigma _l}{P_l} = \left[ {{{\tilde \Sigma }_{l,1}},{{\tilde \Sigma }_{l,2}},  \cdots ,{{\tilde \Sigma }_{l,{\mathcal C}_{\sigma  - 1 + l}^l}} } \right],\]
where $\Sigma$ stands for either $\mathbf X,\mathbf U,\mathbf Y,\mathbf E,\mathbf G$ or $\mathbf W$. Then, Eqs. (\ref{Sylvester_l1})-(\ref{Sylvester_l2}) can be equivalently written as
\begin{equation}\label{tf}
{{\tilde{\mathbf Y}}_{1,\ell }} = \mathcal G_1\left( \lambda_{1,\ell} \right){{\tilde{\mathbf U}}_{1,\ell }},\quad \ell  = 1, \cdots ,\sigma,
\end{equation}
where $\mathcal G_1\left( s \right) = C{\left( {sI - A} \right)^{ - 1}}B + D$. Moreover, Eqs. (\ref{Sylvester_l3})-(\ref{Sylvester_l4}) can be equivalently written as
\begin{equation}\label{tf_2}
{{\tilde{\mathbf Y}}^\prime_{l,\ell }} =  {{\mathcal G}_l}\left( \lambda_{l,\ell} \right){{{\tilde{\mathbf W}}}_{l,\ell }},\quad \ell  = 1, \cdots ,{\mathcal C}_{\sigma  - 1 + l}^l,
\end{equation}
where ${{\tilde{\mathbf Y}}^\prime_{l,\ell }} ={{\tilde{\mathbf Y}}_{l,\ell }} - {{\mathcal G}_1}\left( \lambda_{l,\ell} \right){{\tilde{\mathbf U}}_{l,\ell }} - {{\mathcal H}_{l,\ell }}\left( \lambda_{l,\ell} \right)$,
${{\mathcal G}_l}\left( s \right) = C{\left( {sI - A} \right)^{ - 1}}{{\mathbf{F}}_l} + {{\mathbf{H}}_l}$ and ${{\mathcal H}_{l,\ell }}\left( s \right) = C{\left( {sI - A} \right)^{ - 1}}{{{\tilde{\mathbf E}}}_{l,\ell }} + {{{\tilde{\mathbf G}}}_{l,\ell }}$. Denoting ${{{\tilde{\mathbf X}}}^\prime_{l,\ell }}{\rm{ = }}{\left( {{\lambda _{l,\ell }}{I_n} - A} \right)^{ - 1}}{{\mathbf{F}}_l}$, ${{{{\mathbf X}}}^\prime_l} = \left[ {{{{\tilde{\mathbf X'}}}_{l,1}},{{{\tilde{\mathbf X}}}^\prime_{l,2}}, \cdots ,{{{\tilde{\mathbf X}}}^\prime_{l,{\mathcal C}_{\sigma  - 1 + l}^l}}} \right]P_l^{ - 1}$
and
\begin{equation}\label{mathbf_Zl}
{{\mathbf{Y}}^\prime_l} = \left[ {{{{\tilde{\mathbf Y}}}^\prime_{l,1}},{{{\tilde{\mathbf Y}}}^\prime_{l,2}}, \cdots ,{{{\tilde{\mathbf Y}}}^\prime_{l,{\mathcal C}_{\sigma  - 1 + l}^l}}} \right]P_l^{ - 1},
\end{equation}
Eqs. (\ref{Sylvester_l3})-(\ref{Sylvester_l4}) can be further equivalently rewritten as
\begin{equation}\label{Sylvester_l5}
{{{\mathbf{X}}}_l^\prime}S_1^{\left\langle l \right\rangle } = A{{{\mathbf{X}}}_l^\prime} + {{\mathbf{F}}_l}{{\mathbf{W}}_l},
\end{equation}
\begin{equation}\label{Sylvester_l6}
{{\mathbf{Y}}^\prime_l} = C{{{\mathbf{X}}}_l^\prime} + {{\mathbf{H}}_l}{{\mathbf{W}}_l}.
\end{equation}
Finally, we shall remark that, since the structure of the power-series solution is identical for both the model (\ref{eq_preliminary_nldequ_x0}) and the reduced-order model (\ref{eq_preliminary_nldequ_xi0}), we have presented the results only for (\ref{eq_preliminary_nldequ_xi0}). The corresponding results for (\ref{eq_preliminary_nldequ_x0}) can be easily inferred.

\section{Moment Matching for Linear Dynamics}\label{sec:Moment Matching of Linear Dynamics}

By applying Lemma \ref{lem_stability} to the model (\ref{eq_preliminary_nldequ_x0}), the moment of (\ref{eq_preliminary_nldequ_x0}) at $\left(s\left( \cdot \right),\mathbf u\left( \cdot \right)\right)$ is then parameterized as
\[
{\mathbf{y}^{\rm o}}\left( v \right) = \sum\nolimits_{l = 1}^\infty {{{\mathbf{Y}}_l^{\rm o}}{v^{\left[l\right]}}} ,
\]
where $\mathbf Y_l^{\rm o}$ is known \emph{a priori} for $l\in\mathbb N_{1:\kappa}$. $\mathbf Y_l^{\rm o}$ can be obtained either based on Lemma \ref{lem_stability} given the knowledge of the parameters of model (\ref{eq_preliminary_nldequ_x0}) and the signal generator (\ref{eq_preliminary_nldequ_v0}), or by data-driven approaches \cite{SCARCIOTTI2017Data,Huang2024Identification}. Indeed, $\mathbf Y_l^{\rm o}$, $l\in\mathbb N_{1:\kappa}$ is all we need from the model (\ref{eq_preliminary_nldequ_x0}) in order to achieve moment matching. Based on Definition \ref{def_multiplr_moment_match}, moment matching with precision of degree $\kappa$ naturally requires that
\[
{{\mathbf{Y}}_l^{\rm o}} = {\mathbf{Y}}_l,\quad l\in\mathbb N_{1:\kappa}.
\]
In this section, we focus on matching the moment with first degree precision, that is ${{\mathbf{Y}}_1^{\rm o}} = {\mathbf{Y}}_1$, while the next two sections explore the matching of high-degree terms. 




\subsection{Key Lemmas}\label{subsubsec:KeyLemmas}

Let $\mathcal G_1\left(s\right)=[G_{k,i}\left(s\right)]$, for $k\in\mathbb N_{1:p}$ and $i\in\mathbb N_{1:m}$, where $G_{k,i}\left(s\right)$ is parameterized as
\begin{equation}\label{G_ik}
{G_{k,i}}\left( s \right) = \frac{{{\beta _{k,n_{\rm d},i}}{s^{n_{\rm d}}} + {\beta _{k,n_{\rm d}-1,i}}{s^{n_{\rm d} - 1}} +  \cdots  + {\beta _{k,0,i}}}}{d\left(s\right)}.
\end{equation}
In Eq. (\ref{G_ik}), $d\left(s\right)$ is the monic common denominator of all entries of $\mathcal G_1\left(s\right)$. The degree of $d\left(s\right)$ is denoted by $n_{\rm d}$.

\begin{Lem}[Lemma 6, {\cite{Huang2024Persistence}}]\label{lem-Y=ThetaUR}
Suppose Assumption \ref{asu_spectrumS1} is satisfied. Then there is a nonsingular matrix $R\in\mathbb R^{\sigma\times \sigma}$ such that Eq. (\ref{tf}), or Eqs. (\ref{Sylvester_l1})-(\ref{Sylvester_l2}), is equivalently written as
\begin{equation}\label{Y=ThetaUR}
\mathbf Y_1 = \Theta {{\mathcal U^{\left(n_{\rm d}\right)}}}R,
\end{equation}
where
\[
{{\mathcal{U}^{\left(n_{\rm d}\right)}}} = \left[ {\begin{array}{*{20}{c}}
\mathbf{U}_1\\
{\mathbf{U}_1S_1}\\
 \vdots \\
{\mathbf{U}_1{S_1^{n_{\rm d}}}}
\end{array}} \right]\in\mathbb R^{m\left(1+n_{\rm d}\right)\times \sigma},
\]
and
\[\Theta  = \left[ {\begin{array}{*{20}{c}}
{{\Theta _0}}&{{\Theta _1}}& \cdots &{{\Theta _{n_{\rm d}}}}
\end{array}} \right]\]
with
\[{\Theta _k} = \left[ {\begin{array}{*{20}{c}}
{{\beta _{1,k,1}}}&{{\beta _{1,k,2}}}& \cdots &{{\beta _{1,k,m}}}\\
{{\beta _{2,k,1}}}&{{\beta _{2,k,2}}}& \cdots &{{\beta _{2,k,m}}}\\
 \vdots & \vdots &{}& \vdots \\
{{\beta _{p,k,1}}}&{{\beta _{p,k,2}}}& \cdots &{{\beta _{p,k,m}}}
\end{array}} \right].\]
\end{Lem}

\begin{Rem}
The matrix $R$ depends on $S_1$. According to \cite{Huang2024Persistence}, if $S_1$ is given by
\[\begin{array}{l}
{S_1} = {\rm{blkdiag}}\left( {{\omega _{0,1}}, \cdots ,{\omega _{0,\alpha }},\left[ {\begin{array}{*{20}{c}}
0&{{\omega _1}}\\
{ - {\omega _1}}&0
\end{array}} \right],} \right.
\left. { \cdots ,\left[ {\begin{array}{*{20}{c}}
0&{{\omega _q}}\\
{ - {\omega _q}}&0
\end{array}} \right]} \right),
\end{array}\]
where $\omega _{0,1}=\cdots=\omega _{0,\alpha}=0$ and $\omega_k>0$, for $k=1,2,\cdots,q$, i.e., $\sigma=2q+\alpha$, it follows that
\begin{equation}\label{R}
\begin{array}{l}
R = {\rm{blkdiag}}\left( \frac{1}{{d\left( \omega _{0,1} \right)}},\cdots,\frac{1}{{d\left( \omega _{0,\alpha} \right)}}, \right.\\
\frac{1}{{{{\left| {d\left( {j{\omega _1}} \right)} \right|}^2}}}\left[ {\begin{array}{*{20}{c}}
{{\rm{Re}}\left( {d\left( {j{\omega _1}} \right)} \right)}&{ - {\rm{Im}}\left( {d\left( {j{\omega _1}} \right)} \right)}\\
{{\rm{Im}}\left( {d\left( {j{\omega _1}} \right)} \right)}&{{\rm{Re}}\left( {d\left( {j{\omega _1}} \right)} \right)}
\end{array}} \right],\\
\left. { \cdots ,\frac{1}{{{{\left| {d\left( {j{\omega _q}} \right)} \right|}^2}}}\left[ {\begin{array}{*{20}{c}}
{{\rm{Re}}\left( {d\left( {j{\omega _q}} \right)} \right)}&{ - {\rm{Im}}\left( {d\left( {j{\omega _q}} \right)} \right)}\\
{{\rm{Im}}\left( {d\left( {j{\omega _q}} \right)} \right)}&{{\rm{Re}}\left( {d\left( {j{\omega _q}} \right)} \right)}
\end{array}} \right]} \right).
\end{array}
\end{equation}
$R$ is nonsingular if and only if $A$ and $S_1$ share no common eigenvalues.
\end{Rem}

Of special interest are two cases. Case I: $d\left(s\right)=\det\left(sI_n-A\right)$ and Case II: $d\left(s\right)$ is the \emph{monic least common denominator of all entries} of $\mathcal G_1\left(s\right)$. The two cases are discussed in the sequel.

\subsubsection{Case I}

Let $n_{\rm d}=n$ and
\[\begin{array}{l}
\det \left( {s{I_{n}} - {A}} \right) = {s^{n}} + {a_1}{s^{n - 1}} +  \cdots  + {a_{n}},\\
{\rm{adj}}\left( {s{I_{n}} - {A}} \right) = {A_0}{s^{n - 1}} + {A_1}{s^{n - 2}} +  \cdots  + {A_{n-1}}.
\end{array}\]
Then the following holds.

\begin{Lem}\label{lem_Theta_char_poly}
If $d\left(s\right)=\det \left( {s{I_{n}} - {A}} \right)$, then
\[\Theta  = \left[ {\begin{array}{*{20}{c}}
{{C}}&{{D}}
\end{array}} \right]\mathcal C,\]
where
\[\mathcal C = \left[ {\begin{array}{*{20}{c}}
{{A_{n - 1}}B}& \cdots &{{A_0}B}&{{O_{n \times m}}}\\
{{a_n}{I_m}}& \cdots &{{a_1}{I_m}}&{{I_m}}
\end{array}} \right].\]
\end{Lem}

\begin{Proof}
This can be shown straightforwardly by noting that
\[{\mathcal G_1}\left( s \right) = {C}\frac{{{\rm{adj}}\left( {s{I_{n}} - {A}} \right)}}{{\det \left( {s{I_{n}} - {A}} \right)}}{B} + {D},\]
so that $\Theta_k=CA_{n-1-k}B+a_{n-k}D$, for $k\in\mathbb N_{0:n-1}$ and $\Theta_n=D$.
\end{Proof}

\subsubsection{Case II}

Suppose $d\left(s\right)$ is the \emph{monic least common denominator of all entries} of $\mathcal G_1\left(s\right)$. The degree of $d\left(s\right)$ is denoted by $n_{\rm min}$, i.e., $n_{\rm d}=n_{\rm min}$. If the McMillan degree of $\mathcal G_1\left(s\right)$ is $n_{\rm g}$, then $n_{\rm min}\le n_{\rm g}\le n$.





\subsection{Method I: from $\left(A,B\right)$ to $\left(C,D\right)$}\label{subsec:standard_solution}

The standard method for moment matching of linear dynamics consists in preselecting $\left(A,B\right)$, and then finding $\left(C,D\right)$ that guarantees matching for the selected $\left(A,B\right)$. The result for $D=0$ was presented in \cite{Astolfi2010Model}, but is restated here with a simple proof.

\begin{Lem}[\cite{Astolfi2010Model}]\label{lem-linear_D=0}
Suppose $\left(\mathbf U_1,S_1\right)$ is observable. Then if
\begin{equation}\label{n=sigma}
n\ge \sigma
\end{equation}
there exists a quadruple $\left(A,B,C,D\right)$, with $A\in\mathbb R^{n\times n}$ Hurwitz and $D=0$, such that ${{\mathbf{Y}}_1^{\rm o}} = {\mathbf{Y}}_1$.
\end{Lem}

\begin{Proof}
It suffices to show the existence of $\left(A,B,C,D\right)$ with $D=0$ when $n=\sigma$. Let $A = S_1 - B\mathbf U_1$, where $B$ is such that $A$ is Hurwitz. Such a matrix $B$ exists by the observability of $\left(\mathbf U_1,S_1\right)$. It can be verified that (\ref{Sylvester_l1}) is satisfied with $\mathbf X_1=I_n$. Then, set $C = {\mathbf{Y}}_1^{\rm{o}}$ and $D=0$. It can be verified that (\ref{Sylvester_l2}) is satisfied with ${{\mathbf{Y}}_1}={\mathbf{Y}}_1^{\rm o}$.
\end{Proof}

In fact, the above result can be extended to the case $D\neq0$, given that Assumption \ref{asu_spectrumS1} and the following condition are satisfied.

\begin{Asu}\label{asu_obsv_Method_I}
For $n\ge\sigma-m$, there is a controllable $\left(A,B\right)$ such that $\mathcal C\mathcal U^{\left(n\right)}\in\mathbb R^{\left(n+m\right)\times\sigma}$ has full column rank.
\end{Asu}

Note that ${\rm rank}\left(\mathcal C\right)=n+m$ if and only if $\left(A,B\right)$ is controllable (\cite{Huang2024Persistence}, Lemma 2). Moreover, given that $n=\sigma-m$ and ${\rm rank}\left(\mathbf U_1\right)=m$, it follows that ${\rm rank}\left(\mathcal U^{\left(n\right)}\right)=\sigma$ if and only if $\left(\mathbf U_1,S_1\right)$ is observable (\cite{Chen1999Linear}, Corollary 6.O1). Therefore, Assumption \ref{asu_obsv_Method_I} is actually very mild and it is satisfied generically if ${\rm rank}\left(\mathbf U_1\right)=m$ and $\left(\mathbf U_1,S_1\right)$ is observable. In fact, if $m=1$, it can be shown that Assumption \ref{asu_obsv_Method_I} is equivalent to the observability of $\left(\mathbf U_1,S_1\right)$.

Now suppose that $\left(A,B\right)$ is preselected, and that the solution $\mathbf X_1$ to Eq. (\ref{Sylvester_l1}) is such that
\[\Omega_1  := \left[ {\begin{array}{*{20}{c}}
\mathbf X_1\\
\mathbf U_1
\end{array}} \right]\in\mathbb R^{\left(n+m\right)\times\sigma}\]
has full column rank. $\left(C,D\right)$ can then be obtained based on Eq. (\ref{Sylvester_l2}) by the relation
\begin{equation}\label{CD}
\left[ {\begin{array}{*{20}{c}}
C&D
\end{array}} \right] = {\bf{Y}}_1^{\rm{o}}{\Omega_1 ^{ \dag}},
\end{equation}
if ${\mathbf{Y}}_1={{\mathbf{Y}}_1^{\rm o}}$.

Unlike Lemma \ref{lem-linear_D=0} for $D=0$, in the case of $D\neq0$, Assumption \ref{asu_obsv_Method_I} cannot be relaxed to $\left(\mathbf U_1,S_1\right)$ being observable without imposing the condition ${\rm rank}\left(\mathbf U_1\right)=m$. A counterexample is given by the case in which $n=\sigma-m$, $\left(\mathbf U_1,S_1\right)$ is observable, but $\mathbf U_1$ is rank deficient. In this case $\Omega_1$ is a singular matrix. We state the main result for Method I as follows.

\begin{Thm}\label{lem_full_rankXU}
Suppose Assumptions \ref{asu_spectrumS1} and \ref{asu_obsv_Method_I} are satisfied, and $m<\sigma$. Then if
\begin{equation}\label{n=sigma-m}
n \ge \sigma-m,
\end{equation}
for almost any controllable pair $\left(A,B\right)$, with $A\in\mathbb R^{n\times n}$ Hurwitz, the solution $\mathbf X_1$ to Eq. (\ref{Sylvester_l1}) is such that ${\Omega_1 }$ has full column rank. Consequently, there exists $\left(C,D\right)$ such that the quadruple $\left(A,B,C,D\right)$ satisfies the condition ${\mathbf{Y}}_1={{\mathbf{Y}}_1^{\rm o}}$.
\end{Thm}

\begin{Proof}
Consider Eq. (\ref{Sylvester_l1}) with the output equation:
\begin{equation}\label{Sylvester_2prime}
\Omega_1 = C^\prime{{\mathbf{X}}_1} + D^\prime{{\mathbf{U}}_1},
\end{equation}
where $\left[ {\begin{array}{*{20}{c}}
{{C^\prime }}&{{D^\prime }}
\end{array}} \right] = {I_{m + n}}$. By Lemma \ref{lem-Y=ThetaUR}, Eqs. (\ref{Sylvester_l1}) and (\ref{Sylvester_2prime}) can be written as
\[
\Omega_1 = \Theta^\prime {{\cal U}^{\left( {{n}} \right)}}R,
\]
when $d\left(s\right)=\det \left( {s{I_{n}} - {A}} \right)$, i.e., $n_{\rm d}=n$. By Lemma \ref{lem_Theta_char_poly}, ${\Theta ^\prime } = \left[ {\begin{array}{*{20}{c}}
{{C^\prime }}&{{D^\prime }}
\end{array}} \right]\mathcal C =\mathcal C$. According to Eq. (\ref{R}), $R$ is nonsingular if $d\left(s\right)$ is Hurwitz. Finally by Assumption \ref{asu_obsv_Method_I}, $\Omega_1=\mathcal C\mathcal U^{\left(n\right)}R$ has full column rank if $n\ge\sigma-m$.

Let now $\mathcal S:=\mathbb H^{n\times n}\times \mathbb R^{n\times m}$, where $\mathbb H^{n\times n}\subset\mathbb R^{n\times n}$ is the set of all $n\times n$ Hurwitz matrices. So far we have shown that there exists at least one controllable pair $\left(A,B\right)\in\mathcal S$ such that ${\Omega_1 }$ has full column rank. Since ${\Omega_1 }$ being full column rank and $\left(A,B\right)$ being controllable are both open-set constraints (i.e., both characterized by polynomial inequalities regarding the entries of the pair $\left(A,B\right)$), as long as the intersection of the open sets is non-empty (which has just been proved), it is a dense subset of $\mathcal S$. In other words, for almost any controllable pair $\left(A,B\right)$ with $A$ Hurwitz, ${\Omega_1 }$ has full column rank. This concludes the proof.
\end{Proof}


The pseudo-code of the algorithm for moment matching with precision of degree one based on Theorem \ref{lem_full_rankXU} is given in Algorithm \ref{Alg0}.

\begin{algorithm}
\caption{Moment matching with precision of degree one (if (\ref{n=sigma-m}) holds)}
\label{Alg0}
\textbf{Input:} $\left(\mathbf U_1,S_1,\mathbf Y_1^{\rm o}\right)$\\
\textbf{Output:} $\left(A,B,C,D,\mathbf X_1\right)$
 \begin{enumerate}
    \item Select a controllable pair $\left(A,B\right)$ with $A$ Hurwitz.
    \item Obtain $\mathbf X_1$ by solving the Sylvester equation (\ref{Sylvester_l1}).
    \item Obtain $\left(C,D\right)$ from (\ref{CD}).
\end{enumerate}
\end{algorithm}

\subsection{Method II: from $\left(C,A\right)$ to $\left(B,D\right)$}\label{subsec:Towards_a_lower_order}

In Section \ref{subsec:standard_solution}, a controllable pair \((A, B)\) is preselected first, and then $\left(C,D\right)$ is obtained by solving a linear equation. In this subsection, by contrast, an observable pair \((C, A)\) is preselected, and $\left(B,D\right)$ is subsequently determined by solving a linear equation. It turns out that with this approach we are able to yield lower-order models in many cases. To this end we need the notion of observability index.

\begin{Def}[see e.g. \cite{Chen1999Linear}]\label{def_obsv_index}
The observability index $\mu$ is the smallest integer such that the matrix
\[
{{\mathcal{U}^{\left(\mu-1\right)}}} = \left[ {\begin{array}{*{20}{c}}
\mathbf{U}_1\\
{\mathbf{U}_1S_1}\\
 \vdots \\
{\mathbf{U}_1{S_1^{\mu-1}}}
\end{array}} \right]\in\mathbb R^{m\mu\times \sigma}
\]
satisfies the condition ${\rm rank}\left(\mathcal{U}^{\left(\mu-1\right)}\right)=\sigma$, i.e., $\mathcal{U}^{\left(\mu-1\right)}$ has full column rank.
\end{Def}

\begin{Asu}\label{asu_obsv_index}
$\left(\mathbf U_1,S_1\right)$ has an observability index equal to $\left\lceil {\frac{\sigma }{m}} \right\rceil$, where $\left\lceil a\right\rceil $ denotes the smallest integer that is no less than $a$.
\end{Asu}

\begin{Rem}
It is easy to deduce from Definition \ref{def_obsv_index} that $\mu\ge\left\lceil {\frac{\sigma }{m}} \right\rceil$. Then Assumption \ref{asu_obsv_index} means that $\left(\mathbf U_1,S_1\right)$ has a \textit{minimal} observability index. While Assumption \ref{asu_obsv_index} is a stronger condition compared to observability when $m>1$, it is actually satisfied generically by an observable pair $\left(\mathbf U_1,S_1\right)$. When $m=1$, Assumption \ref{asu_obsv_index} is equivalent to observability. 
\end{Rem}

By eliminating $\mathbf X_1$, Eqs. (\ref{Sylvester_l1})-(\ref{Sylvester_l2}) can be rewritten as
\begin{equation}\label{Psi}
{{\rm vec}\left(\mathbf Y_1\right)} = {\Xi _1}\left[ {\begin{array}{*{20}{c}}
{{\rm{vec}}B}\\
{{\rm{vec}}D}
\end{array}} \right],
\end{equation}
where
\[{\Xi _1} = \left[ {\begin{array}{*{20}{c}}
{\Xi _1^\prime }&{{\bf{U}}_1^{\rm{T}} \otimes {I_p}}
\end{array}} \right]\in {\mathbb R^{p\sigma  \times m\left( {n + p} \right)}},\]

\[\Xi _1^\prime  = \left( {{I_\sigma } \otimes C} \right){\left[ {\left( { - A} \right) \oplus \left( {S_1^{\rm{T}}} \right)} \right]^{ - 1}}\left( {{\bf{U}}_1^{\rm{T}} \otimes {I_n}} \right) \in {\mathbb R^{p\sigma  \times mn}}.\]
As long as the selected $\left(C,A\right)$ is such that $\Xi_1$ has full row rank, $\left(B,D\right)$ can be obtained by the relation
\begin{equation}\label{vecBD}
\left[ {\begin{array}{*{20}{c}}
{{\rm{vec}}B}\\
{{\rm{vec}}D}
\end{array}} \right] = \Xi _1^\dag {\rm{vec}}\left( {{\bf{Y}}_1^{\rm{o}}} \right),
\end{equation}
if ${{\bf{Y}}_1}={{\bf{Y}}_1^{\rm{o}}}$. This approach is stated formally as follows.

\begin{Thm}\label{lem_full_rankU}
Suppose Assumptions \ref{asu_spectrumS1} and \ref{asu_obsv_index} are satisfied, and $m<\sigma$. Then if
\begin{equation}\label{ineq:n}
n \ge p\left( {\left\lceil {\frac{\sigma }{m}} \right\rceil  - 1} \right),
\end{equation}
for almost any observable pair $\left(C,A\right)$, with $A\in\mathbb R^{n\times n}$ Hurwitz, ${\Xi_1 }$ has full row rank. Consequently, there exists $\left(B,D\right)$ such that the quadruple $\left(A,B,C,D\right)$ satisfies ${\mathbf{Y}}_1={{\mathbf{Y}}_1^{\rm o}}$.
\end{Thm}

\begin{Proof}
To show that $\Xi_1$ has full row rank for a given $\left(C,A\right)$, it is equivalent to show that for a given $\left(C,A\right)$ and any $\mathbf Y_1$, there exists a pair $\left(B,D\right)$ satisfying Eq. (\ref{Psi}).

Let $n_{\rm min}= {\left\lceil {\frac{\sigma }{m}} \right\rceil  - 1} $. By Assumption \ref{asu_obsv_index} and Definition \ref{def_obsv_index}, it follows that ${\rm rank}\left(\mathcal U^{\left(n_{\rm min}\right)}\right)=\sigma$. Since $R$ is nonsingular (since $d\left(s\right)$ is Hurwitz) and $\mathcal U^{\left(n_{\rm min}\right)}$ has full column rank, by Lemma \ref{lem-Y=ThetaUR}, for each $\mathbf Y_1\in\mathbb R^{m\times \sigma}$, the equation $\mathbf Y_1 = \Theta {{\mathcal U^{\left(n_{\rm min}\right)}}}R$ has a solution $\Theta$, each of which is associated with a transfer function $\mathcal G_1\left(s\right)$ whose \emph{monic least common denominator of all entries}, i.e., $d\left(s\right)$ (see Section \ref{subsubsec:KeyLemmas}, Case II) has a degree no larger than $n_{\rm min}$. We want to show in the sequel that, if $d\left(s\right)$ is fixed, each $\Theta$ is associated with a common $\left(C,A\right)$.

Based on realization theory, for each $\mathcal G_1\left(s\right)$ associated with $\Theta$, there exist a realization $\left(A,B,C,D\right)$ with $A\in\mathbb R^{n_{\rm min}p\times n_{\rm min}p}$ ($n_{\rm min}p\le n$ according to Eq. (\ref{ineq:n})), moreover, $\left(C,A\right)$ is observable and in canonical form (thus is the same for each $\Theta$). Indeed, one has
\[\begin{array}{l}
A = \left[ {\begin{array}{*{20}{c}}
{ - {a_1}}&1&0& \cdots &0\\
{ - {a_2}}&0&1& \cdots &0\\
 \vdots & \vdots & \vdots & \ddots & \vdots \\
{ - {a_{{n_{\min }} - 1}}}&0&0& \cdots &1\\
{ - {a_{{n_{\min }}}}}&0&0& \cdots &0
\end{array}} \right] \otimes {I_p},\\ C = \left[ {\begin{array}{*{20}{c}}
1&0& \cdots &0
\end{array}} \right] \otimes {I_p},
\end{array}\]
where $d\left(s\right)=s^{n_{\rm min}}+a_1s^{n_{\rm min}-1}+\cdots+a_{n_{\rm min}}$, see e.g. Chapter 4.4 of \cite{Chen1999Linear}. Then, it follows that there also exists a realization $\left(A,B,C,D\right)$, with $A\in\mathbb R^{n\times n}$, corresponding to each transfer function, moreover, $\left(C,A\right)$ is observable and is the same for each $\Theta$.

Let now $\mathcal T:=\mathbb R^{p\times n}\times \mathbb H^{n\times n}$, see the proof of Theorem \ref{lem_full_rankU} for the definition of $\mathbb H^{n\times n}$. So far we have shown that there exists at least one observable pair $\left(C,A\right)\in\mathcal T$ such that ${\Xi_1 }$ has full row rank. Since ${\Xi_1 }$ being full row rank and $\left(C,A\right)$ being observable are both open-set constraints (i.e., both characterized by polynomial inequalities regarding the entries of the pair $\left(C,A\right)$), as long as the intersection of the open sets is non-empty (which has just been proved), it is a dense subset of $\mathcal T$. In other words, for almost any observable pair $\left(C,A\right)$, with $A$ Hurwitz, ${\Xi_1 }$ has full row rank. This concludes the proof.
\end{Proof}

\begin{Rem}
If $m$ is divided by $\sigma$, Eq. (\ref{ineq:n}) becomes $p\sigma \le m\left({n + p}\right)$, which simply means that $\Xi_1$ has no more rows than columns, and thus (\ref{ineq:n}) is necessary that $\Xi$ is full row rank. Otherwise, counterexamples suggest that (\ref{ineq:n}) cannot be strengthened to $p\sigma \le m\left({n + p}\right)$.
\end{Rem}

The pseudo-code of the algorithm for moment matching with precision of degree one based on Theorem \ref{lem_full_rankU} is given in Algorithm \ref{Alg1}.

\begin{algorithm}
\caption{Moment matching with precision of degree one (if (\ref{ineq:n}) holds)}
\label{Alg1}
\textbf{Input:} $\left(\mathbf U_1,S_1,\mathbf Y_1^{\rm o}\right)$\\
\textbf{Output:} $\left(A,B,C,D,\mathbf X_1\right)$
 \begin{enumerate}
    \item Select an observable pair $\left(C,A\right)$ with $A$ being Hurwitz;
    \item Obtain $\left(B,D\right)$ by with (\ref{vecBD});
    \item Obtain $\mathbf X_1$ by solving the Sylvester equation (\ref{Sylvester_l1}).
\end{enumerate}
\end{algorithm}

In some applications, it is desirable that the reduced-order models have $D=0$. In this case Eq. (\ref{Psi}) should be modified to
\begin{equation}\label{Psi_D=0}
{{\rm vec}\left(\mathbf Y_1\right)} = \Xi _1^\prime {\rm{vec}}B.
\end{equation}
The following theorem provides the result for $D=0$.

\begin{Thm}\label{thm-existence2}
Suppose Assumptions \ref{asu_spectrumS1} and \ref{asu_obsv_index} hold, and $m<\sigma$. Then if
\begin{equation}\label{ineq:n_D=0}
n \ge p\left\lceil {\frac{\sigma }{m}} \right\rceil,
\end{equation}
for almost any observable pair $\left(C,A\right)$, with $A$ Hurwitz, ${\Xi^\prime_1 }$ has full row rank and, consequently, there exists $\left(B,D\right)$ such that the quadruple $\left(A,B,C,D\right)$ with $D=0$ satisfies ${\mathbf{Y}}_1={{\mathbf{Y}}_1^{\rm o}}$.
\end{Thm}

\begin{Proof}
Similar to the proof of Theorem \ref{lem_full_rankU}, we need to derive a sufficient condition for the existence of $\Theta$. For $D=0$, one has $\Theta_{n_{\rm d}}=0$. Thus it is sufficient to consider the rank of $\mathcal U^{\left(n_{\rm min}-1\right)}$ instead of $\mathcal U^{\left(n_{\rm min}\right)}$. Replacing $\mathcal U^{\left(n_{\rm min}\right)}$ with $\mathcal U^{\left(n_{\rm min}-1\right)}$ in the proof of Theorem \ref{lem_full_rankU} yields the expected result.
\end{Proof}

\subsection{Comparison of Methods I and II in Linear Moment Matching}

Tables \ref{tab:comparison_Dneq0} and \ref{tab:comparison_D=0} compare which method yields a smaller lower bound for \( n \) under the conditions \( D \neq 0 \) and \( D = 0 \), respectively. It can be observed that when \( p \geq m \), the lower bound obtained by Method I is smaller while, when \( p < m \), Method II may provide a smaller lower bound.

\begin{table*}[ht]
\centering
\caption{Comparison of (\ref{n=sigma-m}) and (\ref{ineq:n}) for $D\neq0$, assuming $m<\sigma$}
\label{tab:comparison_Dneq0}
\begin{tabular}{|c|c|c|}
\hline
\textbf{Condition} & \textbf{Subcondition} & \textbf{Smaller (or equal) Lower Bound for $n$} \\
\hline
\( p < m \) &  \( \sigma \ge \frac{m^2}{m-p} \) &  $n \ge p\left( {\left\lceil {\frac{\sigma }{m}} \right\rceil  - 1} \right),$ \\
\hline
\( p \ge m \) & None &  $n \ge \sigma-m$ \\
\hline
\end{tabular}
\end{table*}

\begin{table*}[ht]
\centering
\caption{Comparison of (\ref{n=sigma}) and (\ref{ineq:n_D=0}) for $D=0$, assuming $m<\sigma$}
\label{tab:comparison_D=0}
\begin{tabular}{|c|c|c|}
\hline
\textbf{Condition} & \textbf{Subcondition} & \textbf{Smaller (or equal) Lower Bound for $n$} \\
\hline
\( p < m \) & \( \sigma \ge \frac{mp}{m-p} \) &  $n \ge p {\left\lceil {\frac{\sigma }{m}} \right\rceil }$ \\
\hline
\( p \ge m \) & None &  $n \ge \sigma$ \\
\hline
\end{tabular}
\end{table*}

\subsection{The $\sigma\le m$ Case}

Methods I and II both focus on the case \( m < \sigma \), which is, of course, the most common scenario. For the sake of completeness and for the subsequent applications (Lemma \ref{lem-existence1} below plays an important role in Section \ref{Moment Matching with the Precision of Degree kappa_infty}), the corresponding results for \( \sigma \leq m \) are also provided.

\begin{Lem}\label{lem-existence1}
Suppose $\mathbf U_1$ has full rank. Assume $\sigma\le m$. Then for any $\left(C,A\right)$
\begin{enumerate}
\item there exists a pair $\left(B,D\right)$ such that ${{\mathbf{Y}}_1^{\rm o}} = {\mathbf{Y}}_1$ and $\mathbf X_1$ can be arbitrarily assigned;
\item there exists a pair $\left(B,D\right)$ with $B=0$ such that ${{\mathbf{Y}}_1^{\rm o}} = {\mathbf{Y}}_1$;
\item if, in addition, $C$ has full row rank, then there exists a pair $\left(B,D\right)$, with $D=0$, such that ${{\mathbf{Y}}_1^{\rm o}} = {\mathbf{Y}}_1$.
\end{enumerate}
\end{Lem}

\begin{Proof}
If $\sigma\le m$, $\mathbf U_1\in\mathbb R^{m\times\sigma}$ must have full column rank. Then,
\begin{enumerate}
\item for any $\left(C,A,\mathbf X_1\right)$, Eqs. (\ref{Sylvester_l1})-(\ref{Sylvester_l2}) with ${{\mathbf{Y}}_1}$ replaced by ${\mathbf{Y}}_1^{\rm o}$ has a solution, for example
\[\left[ {\begin{array}{*{20}{c}}
B\\
D
\end{array}} \right] = \left[ {\begin{array}{*{20}{c}}
{{{\mathbf{X}}_1}{S_1} - A{{\mathbf{X}}_1}}\\
{{{\mathbf{Y}}^{\rm o}_1} - C{{\mathbf{X}}_1}}
\end{array}} \right]{\mathbf{U}}_1^\dag ;\]
\item for any $\left(C,A\right)$, Eqs. (\ref{Sylvester_l1})-(\ref{Sylvester_l2}) with ${{\mathbf{Y}}_1}$ replaced by ${\mathbf{Y}}_1^{\rm o}$ has a solution with $B=0$, for example $\mathbf X_1=0$, $B=0$, and $D = {{\mathbf{Y}}_1^{\rm o}}{\mathbf{U}}_1^\dag$;
\item for any $\left(C,A\right)$ such that $C$ has full row rank, Eqs. (\ref{Sylvester_l1})-(\ref{Sylvester_l2}) with ${{\mathbf{Y}}_1}$ replaced by ${\mathbf{Y}}_1^{\rm o}$ has a solution with $D=0$, for example $\mathbf X_1=C^\dag\mathbf Y_1^{\rm o}$, $B=\left({{\mathbf{X}}_1}S_1- A{{\mathbf{X}}_1}\right){\mathbf{U}}_1^\dag$, and $D = 0$.
\end{enumerate}
\end{Proof}

\section{Nonlinear Moment Matching Based on Method I}\label{Moment Matching with the Precision of Degree kappa_infty}

The objective of this section is to determine $\left(\mathbf F_l,\mathbf H_l\right)$ from Eqs. (\ref{Sylvester_l5})-(\ref{Sylvester_l6}), so that ${{\mathbf{Y}}_l^{\rm o}} = {\mathbf{Y}}_l$ for $l \in\mathbb N_{2:\kappa}$ ($\kappa=\infty$ included). In accordance to Section \ref{subsec:standard_solution}, in this section we assume $n\ge \sigma-m$ for $D\neq0$ and $n\ge \sigma$ for $D=0$. In addition we make the following assumption.


\begin{Asu}\label{asu_rankWl}
$\mathbf W_l$ has full rank for every $l\in\mathbb N_{2:\kappa}$.
\end{Asu}


Note that Eqs. (\ref{Sylvester_l5})-(\ref{Sylvester_l6}) and Eqs. (\ref{Sylvester_l1})-(\ref{Sylvester_l2}) are very alike. Then, according to Lemma \ref{lem-existence1}, given Assumption \ref{asu_rankWl}, to find out whether or not $\left(\mathbf F_l,\mathbf H_l\right)$ has a solution it suffices to compare $\mathcal C_{\sigma-1+l}^l$, which is the dimension of $S_1^{\left\langle l \right\rangle }$, with
\[m_l=\sum\nolimits_{\scriptstyle i + r = l,\hfill\atop
\scriptstyle i,r \in\mathbb N\hfill} {\mathcal C_{n -1 + i}^i\mathcal C_{m - 1 + r}^r},\]
which is the number of input channels of $\left(A,\mathbf F_l,C,\mathbf H_l\right)$. To this end, we have the following result.

\begin{Lem}\label{lem-equality}
If $n=\sigma-m$, then
\begin{equation}\label{Inductive_neq}
\mathcal C_{\sigma  - 1 + l}^l= m_l, \quad l\in\mathbb N_2.
\end{equation}
\end{Lem}

\begin{Proof}
See Appendix \ref{app_lem-equality}.
\end{Proof}


\begin{Cor}\label{cor-inequality}
If $n\ge \sigma- m$, then $\mathcal C_{\sigma  - 1 + l}^l\le m_l$, $l\in\mathbb N_2$.
\end{Cor}

\begin{Proof}
This is a direct consequence of Lemma \ref{lem-equality}, since $\mathcal C_{\sigma  - 1 + l}^l$ is a monotonically increasing function of $\sigma$.
\end{Proof}


We now give the main result of this section.

\begin{Thm}\label{thm-MM_lth}
Suppose Assumptions \ref{asu_CM}, \ref{asu_obs}, \ref{asu_spectrumS1}, \ref{asu_power_series_converge}, \ref{asu_obsv_Method_I} and \ref{asu_rankWl} are satisfied. Suppose (\ref{n=sigma-m}) holds. Then there exists a quadruple $\left(A,B,C,D\right)$, with $A\in\mathbb R^{n\times n}$ Hurwitz, such that ${{\mathbf{Y}}_1^{\rm o}} = {\mathbf{Y}}_1$, and
\begin{enumerate}
\item there exists a pair $\left(\mathbf F_l,\mathbf H_l\right)$ such that ${{\mathbf{Y}}_l^{\rm o}} = {\mathbf{Y}}_l$ for $l\in\mathbb N_{2:\kappa}$ ($\kappa=\infty$ included) and $\mathbf X_l$ can be arbitrarily assigned;
\item there exists a pair $\left(\mathbf F_l,\mathbf H_l\right)$, with $\mathbf F_l=0$, such that ${{\mathbf{Y}}_l^{\rm o}} = {\mathbf{Y}}_l$ for $l\in\mathbb N_{2:\kappa}$ ($\kappa=\infty$ included);
\item if, in addition, $C$ has full row rank, there exists a pair $\left(\mathbf F_l,\mathbf H_l\right)$, with $\mathbf H_l=0$, such that ${{\mathbf{Y}}_l^{\rm o}} = {\mathbf{Y}}_l$ for $l\in\mathbb N_{2:\kappa}$ ($\kappa=\infty$ included).
\end{enumerate}
\end{Thm}

\begin{Proof}
Based on Section \ref{subsec:standard_solution}, there exists a quadruple $\left(A,B,C,D\right)$ such that ${{\mathbf{Y}}_1^{\rm o}} = {\mathbf{Y}}_1$. Note that $\left(C,A\right)$ is inherited by $\mathcal G_l\left(s\right)$. Then according to Corollary \ref{cor-inequality}, it follows that $\mathcal C_{\sigma  - 1 + l}^l\le m_l$ for every $l\in\mathbb N_2$, which, according to Lemma \ref{lem-existence1}, implies 1), 2) and 3). This concludes the proof.
\end{Proof}

\begin{Thm}\label{thm-MM_lth_D=0}
Suppose Assumptions \ref{asu_CM}, \ref{asu_obs}, \ref{asu_spectrumS1}, \ref{asu_power_series_converge}, \ref{asu_obsv_Method_I} and \ref{asu_rankWl} are satisfied. Suppose (\ref{n=sigma}) holds. Then there exists a quadruple $\left(A,B,C,D\right)$, with $A\in\mathbb R^{n\times n}$ Hurwitz and $D=0$, such that ${{\mathbf{Y}}_1^{\rm o}} = {\mathbf{Y}}_1$, and 1), 2) and 3) of Theorem \ref{thm-MM_lth} are satisfied.
\end{Thm}

\begin{Proof}
The proof can be carried out similarly to that of Theorem \ref{thm-MM_lth}.
\end{Proof}

If the assumptions of Theorem \ref{thm-MM_lth} are all satisfied, the algorithm for moment matching to  $\kappa$-th-degree precision based on Theorem \ref{thm-MM_lth} is given in pseudo-code in Algorithm \ref{Alg2}.

\begin{algorithm}
\caption{Moment Matching with the Precision of Degree $\kappa$ (if (\ref{n=sigma-m}) holds)}
\label{Alg2}
\textbf{Input:} $\left(A,B,C,D,\mathbf X_1\right)$ and $\left(\mathbf U_l,S_l,\mathbf Y_l^{\rm o}\right)$, $l\in\mathbb N_{1:\kappa}$;\\
\textbf{Output:} $\left(\mathbf F_l,\mathbf H_l\right)$, $l\in\mathbb N_{2:\kappa}$.
 \begin{enumerate}
   \item \textbf{For} $s=2:\kappa$\\
          \textbf{Step $1$}:\\
           Compute $\mathbf E_s$, $\mathbf G_s$ and $\mathbf W_s$ by Eqs. (\ref{Wl}), (\ref{Wir}), (\ref{El}) and (\ref{Gl});\\
          \textbf{Step $2$}:\\
           Solve $\left(\mathbf F_s,\mathbf H_s,\mathbf X_s\right)$ by solving the linear equations (\ref{Sylvester_l3})-(\ref{Sylvester_l4}), where ${ {\mathbf{Y}}_{s}}={\mathbf{Y}}_{s}^{\rm o}$; (Note: the users have the freedom to specify the value of either $\mathbf X_s$ or $\mathbf F_s$, or specify the value of $\mathbf H_s$ if $C$ has full row rank)\\
          \textbf{End For}
\end{enumerate}
\end{algorithm}

\section{Nonlinear Moment Matching Based on Method II}\label{Moment Matching with the Precision of Degree kappa}

The objective of this section is to determine  $\left(\mathbf F_l,\mathbf H_l\right)$ from Eqs. (\ref{Sylvester_l5})-(\ref{Sylvester_l6}), so that ${{\mathbf{Y}}_l^{\rm o}} = {\mathbf{Y}}_l$, for $l\in\mathbb N_{2:\kappa}$. In accordance to Section \ref{subsec:Towards_a_lower_order}, we assume that (\ref{ineq:n}) holds for $D\neq0$ and (\ref{ineq:n_D=0}) holds for $D=0$. In addition we make the following assumption.

\begin{Asu}\label{asu_WlS}
$\left(\mathbf W_l,{S_1^{\left\langle l \right\rangle }}\right)$ has an observability index equal to $\left\lceil {\frac{{{\cal C}_{\sigma  - 1 + l}^l}}{{{m_l}}}} \right\rceil$ for every $l\in\mathbb N_{2:\kappa}$.
\end{Asu}

Analogously to Eq. (\ref{Psi}) for $D\neq0$ and Eq. (\ref{Psi_D=0}) for $D=0$, we can rewrite Eqs. (\ref{Sylvester_l5})-(\ref{Sylvester_l6}) in the form
\begin{equation}\label{calML}
{\rm vec}\left(\mathbf Y^\prime_l\right) = {\Xi_l}\left[ {\begin{array}{*{20}{c}}
{{\rm{vec}}{{\mathbf{F}}_l}}\\
{{\rm{vec}}{{\mathbf{H}}_l}}
\end{array}} \right],
\end{equation}
if ${\mathbf{H}}_l\neq0$, or in the form
\begin{equation}\label{calML_D=0}
{\rm vec}\left(\mathbf Y^\prime_l\right) = \Xi^\prime_l{{\rm{vec}}{\mathbf{F}}_l},
\end{equation}
if ${\mathbf{H}}_l=0$, where
\[{\Xi _l} = \left[ {\begin{array}{*{20}{c}}
{\Xi _l^\prime }&{{\bf{W}}_l^{\rm{T}} \otimes {I_p}}
\end{array}} \right]\in {\mathbb R^{p{\cal C}_{\sigma  - 1 + l}^l \times {m_l}\left( {n + p} \right)}},\]
and
\[\begin{array}{l}
\Xi _l^\prime  = \left( {{I_{\mathcal C_{\sigma  - 1 + l}^l}} \otimes C} \right){\left[ {\left( { - A} \right) \oplus {{\left( {S_1^{\left\langle l \right\rangle }} \right)}^{\rm{T}}}} \right]^{ - 1}}\left( {{\bf{W}}_l^{\rm{T}} \otimes {I_n}} \right)
 \in\mathbb R^{p\mathcal C_{\sigma  - 1 + l}^l \times m_ln} .
\end{array}\]
The main result of this section is now given.

\begin{Thm}\label{thm-MM_lth_general}
Suppose Assumptions \ref{asu_CM}, \ref{asu_obs}, \ref{asu_spectrumS1}, \ref{asu_power_series_converge}, \ref{asu_obsv_index} and \ref{asu_WlS} are satisfied. Assume Eq. (\ref{ineq:n}) and the condition
\begin{equation}\label{Inductive_ineq_kappa}
n \ge p \left(\left\lceil {\frac{{{\cal C}_{\sigma  -1+ \kappa}^\kappa}}{{{\cal C}_{m+n  -1+ \kappa}^\kappa}}} \right\rceil  - 1\right)
\end{equation}
hold. Then there exist a quadruple $\left(A,B,C,D\right)$, with $A\in\mathbb R^{n\times n}$ Hurwitz, and $\left(\mathbf F_l,\mathbf H_l\right)$ for $l\in\mathbb N_{2:\kappa}$, such that ${\mathbf{Y}}_l={{\mathbf{Y}}_l^{\rm o}}$ for $l\in\mathbb N_{1:\kappa}$.
\end{Thm}

\begin{Proof}
Similar to Theorem \ref{lem_full_rankU}, a lower bound for $n$ can be obtained from Eqs. (\ref{Sylvester_l5})-(\ref{Sylvester_l6}) as
\begin{equation}\label{Inductive_ineq_l}
n \ge p \left(\left\lceil {\frac{{{\cal C}_{\sigma  - 1 + l}^l}}{{{m_l}}}} \right\rceil  - 1\right) ,\quad l\in\mathbb N_{2:\kappa},
\end{equation}
where $m_l=\mathcal C_{n+m  - 1 + l}^l$ by Lemma \ref{lem-equality}. By Corollary \ref{cor-inequality}, we know that Eq. (\ref{Inductive_ineq_kappa}) and (\ref{Inductive_ineq_l}) hold trivially if $n\ge\sigma- m$. If $n<\sigma- m$, (\ref{Inductive_ineq_kappa}) and (\ref{Inductive_ineq_l}) are equivalent, this is because ${\frac{{{\cal C}_{\sigma  - 1 + l}^l}}{{{m_l}}}}$ is monotonically increasing with respect to $l$.
\end{Proof}

A simple estimate of the right-hand side of (\ref{Inductive_ineq_kappa}) is given in the next statement.

\begin{Cor}\label{cor_inequality_l_th}
Eq. (\ref{Inductive_ineq_kappa}) holds if
\begin{equation}\label{Inductive_eq_l}
n \ge p \left(\left\lceil {{{\left( {\frac{\sigma }{{m + n}}} \right)}^\kappa }} \right\rceil  - 1\right) .
\end{equation}
\end{Cor}

\begin{Proof}
Eq. (\ref{Inductive_ineq_l}) and (\ref{Inductive_eq_l}) hold trivially if $n\ge\sigma- m$. If $n<\sigma- m$, Eq. (\ref{Inductive_eq_l}) implies (\ref{Inductive_ineq_kappa}) since
\[\frac{{C_{\sigma  - 1 + \kappa }^\kappa }}{{C_{m + n - 1 + \kappa }^\kappa }} = \frac{{\prod\nolimits_{l = 1}^\kappa  {\sigma  - 1 + l} }}{{\prod\nolimits_{l = 1}^\kappa  {m + n - 1 + l} }} \le {\left( {\frac{\sigma }{{m + n}}} \right)^\kappa }.\]
\end{Proof}

\begin{Thm}\label{thm-MM_lth_D=0_general}
Suppose that Assumptions \ref{asu_CM}, \ref{asu_obs}, \ref{asu_spectrumS1}, \ref{asu_power_series_converge}, \ref{asu_obsv_index} and \ref{asu_WlS} are satisfied. If both Eq. (\ref{ineq:n_D=0}) and the condition
\begin{equation}\label{Inductive_ineq_D=0_kappa}
n \ge p {\left\lceil {\frac{{{\cal C}_{\sigma  -1 + \kappa}^\kappa}}{{{\cal C}_{m+n +1 - \kappa }^\kappa}}} \right\rceil }
\end{equation}
hold, then there exist a quadruple $\left(A,B,C,D\right)$, with $A\in\mathbb R^{n\times n}$ Hurwitz, $D=0$, and $\left(\mathbf F_l,\mathbf H_l\right)$, with $\mathbf H_l=0$ for $l\in\mathbb N_{2:\kappa}$, such that ${\mathbf{Y}}_l={{\mathbf{Y}}_l^{\rm o}}$ for $l\in\mathbb N_{1:\kappa}$.
\end{Thm}

\begin{Proof}
This result can be proven analogously to Theorem \ref{thm-MM_lth_general}.
\end{Proof}

If the assumptions of Theorem \ref{thm-MM_lth_general} are all satisfied, the algorithm for moment matching with precision of degree $\kappa$ based on Theorem \ref{thm-MM_lth_general}  is given in pseudo-code in Algorithm \ref{Alg3}.

\begin{algorithm}
\caption{Moment Matching with Precision of Degree $\kappa$}
\label{Alg3}
\textbf{Input:} $\left(A,B,C,D,\mathbf X_1\right)$ and $\left(\mathbf U_l,S_l,\mathbf Y_l^{\rm o}\right)$, $l\in\mathbb N_{1:\kappa}$;\\
\textbf{Output:} $\left(\mathbf F_l,\mathbf H_l,\mathbf X_l\right)$, $l\in\mathbb N_{2:\kappa}$.
 \begin{enumerate}
 \item \textbf{For} $s=2:\kappa$\\
          \textbf{Step $1$}: \\
          Compute $\mathbf E_s$, $\mathbf G_s$ and $\mathbf W_s$ by Eqs. (\ref{Wl}), (\ref{Wir}), (\ref{El}) and (\ref{Gl});\\
          \textbf{Step $2$}: \\
          Obtain $\mathbf Y^\prime_s$ with Eqs. (\ref{tf_2})-(\ref{mathbf_Zl}), where ${ {\mathbf{Y}}_{s}}={\mathbf{Y}}_{s}^{\rm o}$;\\
          \textbf{Step $3$}: \\
          Solve $\left(\mathbf F_s,\mathbf H_s\right)$ from the linear equations (\ref{calML});\\
          \textbf{Step $4$}: \\
          Obtain $\mathbf X_s$ by solving the Sylvester equation (\ref{Sylvester_l3});\\
          \textbf{End For}
\end{enumerate}
\end{algorithm}

\subsection{Comparison of Methods I and II in Nonlinear Moment Matching}




When $p\ge m$, Method I can obtain a lower bound of model order that is smaller than Method II. When $p< m$, Method II may sometimes achieve a smaller lower bound of model order than Method I. However, due to the complexity of the specific conditions, we will not provide theoretical results here. In the next section, we will demonstrate this point through numerical examples.

In addition, in Eq. (\ref{n=sigma}) of Theorem \ref{thm-MM_lth_D=0}, the value of $n$ is independent of $\kappa$, which means that $\kappa$ can be as large as desired without increasing $n$. In Eq. (\ref{Inductive_ineq_D=0_kappa}) of Theorem \ref{thm-MM_lth_general}, however, $\kappa$ is upper bounded given any fixed $n$. Moreover, Theorem \ref{thm-MM_lth_general} requires that Assumption \ref{asu_WlS} holds while Theorem \ref{thm-MM_lth_D=0} requires that Assumption \ref{asu_rankWl} holds, which is a weaker condition compared to Assumption \ref{asu_WlS}. These are some advantages of Method I over Method II.

\section{Numerical Examples}\label{sec:Numerical Examples}

In this section the proposed model reduction methods are verified through numerical examples. The section is divided into two parts depending on whether $D=0$ in the reduced-order models. 

\subsection{Case I: $D\neq0$}

We consider a polynomial nonlinear system with $n^{\rm o}=6$ and $L=3$. The parameters of the linear dynamics are given by
\[\begin{array}{l}
{A^{\rm{o}}} = {\rm{blkdiag}}\left( { - 0.5,\left[ {\begin{array}{*{20}{c}}
{ - 1}&{0.2}\\
{ - 0.2}&{ - 1}
\end{array}} \right],\left[ {\begin{array}{*{20}{c}}
{ - 1.5}&{0.1}\\
{ - 0.1}&{ - 1.5}
\end{array}} \right], - 2} \right),\\
{B^{\rm{o}}} = {\left[ {\begin{array}{*{20}{c}}
1&1&0&{ - 1}&0&1\\
{ - 2}&0&1&{ - 1}&{ - 1}&{ - 1}
\end{array}} \right]^{\rm{T}}},\\
{C^{\rm{o}}} = \left[ {\begin{array}{*{20}{c}}
{1.3}&1&0&0&1&{ - 2}
\end{array}} \right],\\
{D^{\rm{o}}} = \left[ {\begin{array}{*{20}{c}}
0&0
\end{array}} \right],
\end{array}\]
that is $m=2$ and $p=1$. Moreover, $\mathbf F^{\rm o}_2$ is a $6\times36$ matrix, the entries of which are all zeros except $\mathbf F^{\rm o}_2\left(1,2\right)=-1$, $\mathbf F^{\rm o}_2\left(2,5\right)=2$ and $\mathbf F^{\rm o}_2\left(3,3\right)=0.5$; $\mathbf H^{\rm o}_2$ is a $1\times36$ matrix, the entries of which are all zeros except $\mathbf H^{\rm o}_2\left(1,6\right)=3$.

The reduced-order models are to match the moment of the original system at (\ref{power_series3}), with the non-zero matrices given by
\begin{equation}\label{exo_case_I}
\begin{array}{l}
{S_1} = {\rm{blkdiag}}\left( {0,\left[ {\begin{array}{*{20}{c}}
0&{0.5}\\
{ - 0.5}&0
\end{array}} \right],\left[ {\begin{array}{*{20}{c}}
0&1\\
{ - 1}&0
\end{array}} \right]} \right),\\
{\mathbf U_1} = \left[ {\begin{array}{*{20}{c}}
1&1&0&1&0\\
{ - 1}&0&1&0&1
\end{array}} \right].\\
\end{array}
\end{equation}
Then we have $\sigma=5$, and moreover, $\mathbf U_1$ is full rank, and $\left(\mathbf U_1,S_1\right)$ is observable. By Lemma \ref{lem_stability}, the moment of the original system is obtained up to degree $l=4$, that is,
\begin{equation}\label{yo2}
{{\mathbf{y}}^{\rm{o}}}\left( v \right) = \sum\limits_{l = 1}^4 {{\mathbf{Y}}_l^{\rm{o}}{v^{\left[ l \right]}}}  + o\left( {{{\left\| v \right\|}^4}} \right),
\end{equation}
where
\[\begin{array}{l}
{\bf{Y}}_1^{\rm{o}} = \left[ {\begin{array}{*{20}{c}}
{7.23}&{ - 1.23}&{ - 3.59}
\end{array}\begin{array}{*{20}{c}}
{ - 1.62}&{ - 1.85}
\end{array}} \right],\\
{\bf{Y}}_2^{\rm{o}} = \left[ {\begin{array}{*{20}{c}}
{12.99}&{ - 3.50}&{ - 15.68}&{ - 3.11}&{ - 12.47}&{2.12}
\end{array}} \right.\\
\begin{array}{*{20}{c}}
{5.71 \times {{10}^{ - 1}}}&{1.33}&{ - 1.14}&{5.74}&{2.54}&{7.87}
\end{array}\\
\left. {\begin{array}{*{20}{c}}
{7.14 \times {{10}^{ - 1}}}&{1.11}&{3.25}
\end{array}} \right]
\end{array}\]
and $\mathbf Y^{\rm o}_3\in\mathbb R^{1\times 35}$, $\mathbf Y^{\rm o}_4\in\mathbb R^{1\times 70}$ are non-zero matrices, the entries of which are omitted due to space limitation.

\subsubsection{Method I}

According to Theorem \ref{thm-MM_lth}, there exists a reduced-order model of dimension $n=\sigma-m=3$ that matches the moment with precision of degree $\kappa=\infty$. In this simulation, we compute the first four degrees, i.e., $\kappa=4$. We select $\lambda_1=-0.5,\lambda_2=-1+0.2j,\lambda_3=-1-0.2j$ to be the eigenvalues of $A$. Running Algorithms \ref{Alg0} and \ref{Alg2} up to $\kappa=4$ gives the model with $\mathbf F_l=0$ for $l=2,3,4$ (i.e., it has a linear state-equation), and
\[\begin{array}{l}
A =
\begin{bmatrix}
-0.50 & 0 & 0 \\
0 & -1.00 & 0.20 \\
0 & -0.20 & -1.00
\end{bmatrix}
,\quad B =
\begin{bmatrix}
1.00 & 0 \\
0 & 1.00 \\
1.00 & 0
\end{bmatrix}
,\\
C =
\begin{bmatrix}
-1.81  & -8.47  & 4.92
\end{bmatrix}
,\quad D =
\begin{bmatrix}
1.26  & 2.60
\end{bmatrix}
,
\end{array}\]
moreover, $\mathbf H_2=[{\mathbf{H}}_{2,0},{\mathbf{H}}_{1,1},{\mathbf{H}}_{0,2}]$ where
\[\begin{array}{l}
{{\bf{H}}_{2,0}} = \left[ {3.60 \times {{10}^1}\;4.23 \times {{10}^1}\; - 1.26 \times {{10}^2}} \right],\\
{{\bf{H}}_{1,1}} = \left[ { - 1.78 \times {{10}^{ - 1}}\; - 2.16 \times {{10}^1}\; - 1.29 \times {{10}^1}\; - 8.43} \right.\\
\quad\quad\quad\quad\quad\quad\left. { - 1.79 \times {{10}^1}\;3.72 \times {{10}^1}} \right],\\
{{\bf{H}}_{0,2}} = \left[ {6.18\; - 2.14\;2.30} \right].
\end{array}\]
and $\mathbf H_3\in\mathbb R^{1\times 35}$, $\mathbf H_4\in\mathbb R^{1\times 70}$ are non-zero matrices, the entries of which are omitted due to space limitation.

\subsubsection{Method II}
According to Theorem \ref{thm-MM_lth_general}, there exists a reduced-order model of dimension $n =2$ that matches the moment of the original model at (\ref{power_series3}) with precision of degree $\kappa=2$ (unfortunately, Assumption \ref{asu_WlS} does not hold for $\kappa=3$). To obtain the model, we select $\lambda_1=-1+0.2j,\lambda_2=-1-0.2j$ as the eigenvalues of $A$. Running Algorithms \ref{Alg1} and \ref{Alg3} up to $\kappa=2$ gives
\[\begin{array}{l}
A = \begin{bmatrix}
-1.00 & 0.20 \\
-0.20 & -1.00
\end{bmatrix}
,\quad
B =
\begin{bmatrix}
1.00 \times 10^{1} & 2.21  \\
-5.18  & -8.12
\end{bmatrix}
,\\
C = \begin{bmatrix}
1.00 & 1.00
\end{bmatrix},\quad
D = \begin{bmatrix}
-1.34 \times 10^{-1} & 2.05
\end{bmatrix},
\end{array}\]
moreover,  $\mathbf F_2=[{\mathbf{F}}_{2,0},{\mathbf{F}}_{1,1},{\mathbf{F}}_{0,2}]$ where
\[\begin{array}{l}
{{\mathbf{F}}_{2,0}} = \begin{bmatrix}
1.04  & 1.54  & 3.59 \times 10^{-1} \\
-7.00 \times 10^{-1} & -8.15 \times 10^{-1} & -5.53 \times 10^{-2}
\end{bmatrix},\\
{{\mathbf{F}}_{1,1}} = \begin{bmatrix}
0.00 & 0.00 & 0.00 & 0.00 \\
0.00 & -2.68 \times 10^{-1} & 0.00 & 0.00
\end{bmatrix},\\
{{\mathbf{F}}_{0,2}} = O_{2\times 3},
\end{array}\]
and $\mathbf H_2=[{\mathbf{H}}_{2,0},{\mathbf{H}}_{1,1},{\mathbf{H}}_{0,2}]$ where
\[
\begin{array}{l}
{\mathbf{H}}_{2,0} = \left[
1.20 \times 10^{-1} \; -2.67 \times 10^{-2} \; 1.27 \times 10^{-2}
\right], \\
{\mathbf{H}}_{1,1} = \left[ 
-6.01 \times 10^{-1} \; -5.76 \times 10^{-2} \;7.30 \; -2.76 \times 10^{-1}
\right], \\
{\mathbf{H}}_{0,2} = \left[
3.41 \; 0.00 \; 0.00
\right].
\end{array}
\]

\begin{figure*}
    \centering
    \includegraphics[width=0.8\textwidth]{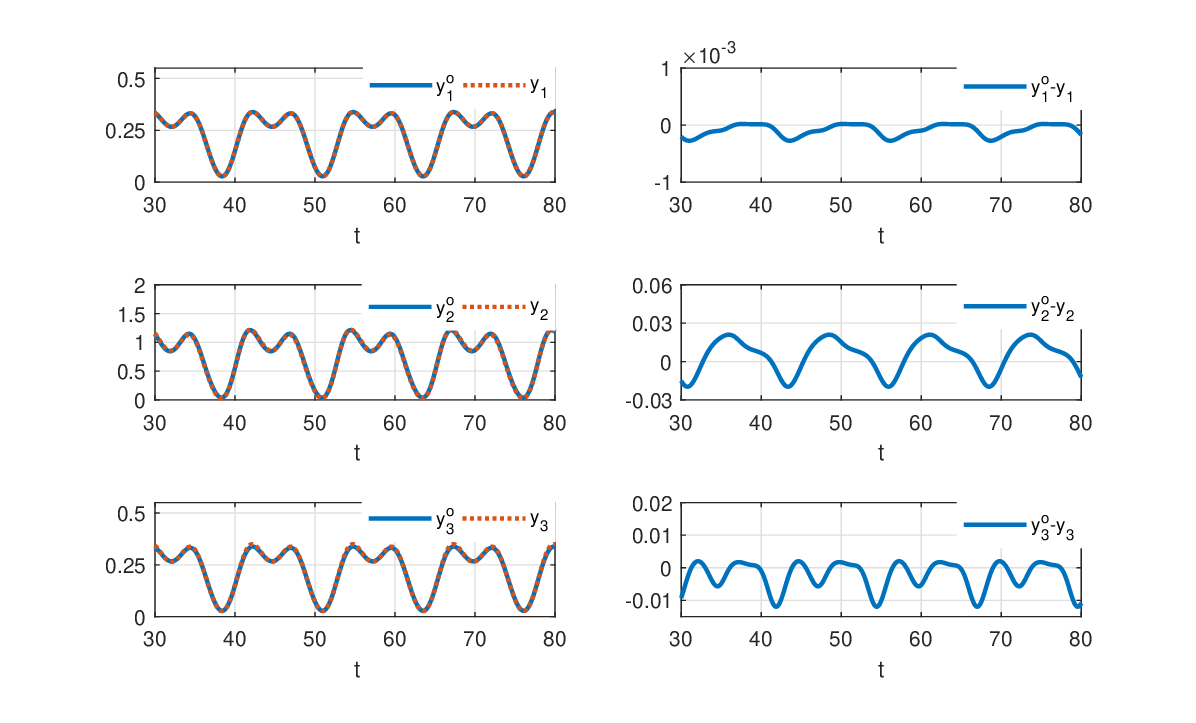}
    \caption{Method I with $D\neq0$: steady-state response $y^{\rm o}$ of the original model versus $y$ of the reduced model. The upper figures show the steady-state responses by Method I with initial condition $v\left(0\right)=0.03\times [1,0,1,0,1]^{\rm T}$ (left) and the error between the responses (right); The middle figures show the steady-state responses by Method I with initial condition $v\left(0\right)=0.1\times [1,0,1,0,1]^{\rm T}$ (left) and the error between the responses (right); the bottom figures show the steady-state responses by Method II with initial condition $v\left(0\right)=0.03\times [1,0,1,0,1]^{\rm T}$ (left) and the error between the responses (right).}
    \label{fig_1}
\end{figure*}

Simulation results are shown in Figure \ref{fig_1}, where both the output of the original system and of the reduced-order model excited by the exo-system (\ref{exo_case_I}) are plotted. The following points can be observed from the simulation.

\begin{enumerate}
    \item For Method I, when the initial state $v(0)$ increases from $0.03 \times [1,0,1,0,1]^{\rm T}$ to $0.1 \times [1,0,1,0,1]^{\rm T}$, the peak of the steady-state output error increases by approximately 100 times. This indicates that the reduced-order model fits small signals much better than large signals. For Method II, a similar phenomenon is observed. Due to the large fitting error when $v(0) = 0.1 \times [1,0,1,0,1]^{\rm T}$, the corresponding simulation results are omitted for Method II. This phenomenon may arises partly due to the truncation effect caused by the finite $\kappa$, and partly because the $\mathscr{C}^\infty$ center manifold of interest in this paper only exists near the equilibrium point.
    
    \item For the same initial state $v(0) = 0.03 \times [1,0,1,0,1]^{\rm T}$, Method I performs better than Method II. This is because Method I uses $\kappa = 4$, while Method II uses $\kappa = 2$. On the other hand, Method I generates a third-order model, while Method II generates a second-order model, which is a lower-order one.
\end{enumerate}

\subsection{Case II: $D=0$}

In the case $D=0$, we consider the model reduction problem of the RL Ladder circuit with nonlinear resistors:
\begin{equation}\label{RL_Ladder}
\begin{array}{l}
{{\dot x}_1} = {u_1} - 2\left( {{x_1} + {u_2}} \right) + {x_2} - {{x_1^2} \mathord{\left/
 {\vphantom {{x_1^2} 2}} \right.
 \kern-\nulldelimiterspace} 2} - {{x_1^3} \mathord{\left/
 {\vphantom {{x_1^3} 3}} \right.
 \kern-\nulldelimiterspace} 3},\\
{{\dot x}_2} = \left( {{x_1} + {u_2}} \right) - 2{x_2} + {x_3} - {{x_2^2} \mathord{\left/
 {\vphantom {{x_2^2} 2}} \right.
 \kern-\nulldelimiterspace} 2} - {{x_2^3} \mathord{\left/
 {\vphantom {{x_2^3} 3}} \right.
 \kern-\nulldelimiterspace} 3},\\
\begin{array}{*{20}{c}}
{{{\dot x}_i} = {x_{i - 1}} - 2{x_i} + {x_{i + 1}} - {{x_i^2} \mathord{\left/
 {\vphantom {{x_i^2} 2}} \right.
 \kern-\nulldelimiterspace} 2} - {{x_i^3} \mathord{\left/
 {\vphantom {{x_i^3} 3}} \right.
 \kern-\nulldelimiterspace} 3},}&{i = 3,4, \cdots ,99,}
\end{array}\\
{{\dot x}_{100}} = {x_{99}} - 2{x_{100}} - {{x_{100}^2} \mathord{\left/
 {\vphantom {{x_{100}^2} 2}} \right.
 \kern-\nulldelimiterspace} 2} - {{x_{100}^3} \mathord{\left/
 {\vphantom {{x_{100}^3} 3}} \right.
 \kern-\nulldelimiterspace} 3},\\
y = {x_1},
\end{array}
\end{equation}
where $x\left(t\right)=[x_1\left(t\right),\cdots,x_{100}]^{\rm T\left(t\right)}\in\mathbb R^{100}$ is the state, $u\left(t\right)=[u_1\left(t\right),u_2\left(t\right)]^{\rm T}\in\mathbb R^{2}$ is the input, $y\left(t\right)\in\mathbb R$ is the output. This model differs from the one in \cite{KAWANO2021Empirical} only in that this model has two voltage sources, $u_1$ and $u_2$, whereas the one in \cite{KAWANO2021Empirical} has one, i.e., $u_1$. It is observed that $n^{\rm o}=100$, $m=2$, $p=1$ and $L=3$.

The reduced-order model is to match the moment of the original system at (\ref{power_series3}), with the non-zero matrices given by
\begin{equation}\label{exo_case_II}
\begin{array}{l}
{S_1} = {\rm{blkdiag}}\left( {\left[ {\begin{array}{*{20}{c}}
0&{100\pi t}\\
{ - 100\pi t}&0
\end{array}} \right],\left[ {\begin{array}{*{20}{c}}
0&{200\pi t}\\
{ - 200\pi t}&0
\end{array}} \right]} \right),\\
{{\bf{U}}_1} = \left[ {\begin{array}{*{20}{c}}
0&1&0&1\\
1&0&1&0
\end{array}} \right],\\
{{\bf{U}}_2} = \left[ {\begin{array}{*{20}{c}}
0&0&1&0&0&0&0&0&0&0\\
0&0&0&0&0&0&1&0&0&0
\end{array}} \right].
\end{array}
\end{equation}
Then we have $\sigma = 4$ and moreover, $\mathbf U_1$ is full rank, and $\left(\mathbf U_1, S_1\right)$ is observable. Due to the high order of the system (\ref{RL_Ladder}), there are computational difficulties when calculating the coefficients of the moments (i.e., $\mathbf{Y}^{\rm o}_l$, $l=1,\cdots,\kappa$) using Lemma \ref{lem_stability}, particularly because the dimension of $M^{100}_i$ is enormous. Therefore, here we employ a data-driven approach to compute $\mathbf{Y}^{\rm o}_l$. The moment of the original system up to degree $\kappa=3$ is obtained from input-output data with the method described in \cite{Huang2024Identification}. The entries of $\mathbf Y_1^{\rm o}$, $\mathbf Y_2^{\rm o}$ and $\mathbf Y_3^{\rm o}$ are omitted due to space limitations.

\subsubsection{Method I}

According to Theorem \ref{thm-MM_lth}, there exists a reduced-order model of dimension $n=\sigma=4$ that matches the moment with precision $\kappa=\infty$. In this simulation, we compute the first three degrees, i.e., $\kappa=3$. We select $\lambda_k=-k$, $k=1,2,3,4$, to be the eigenvalues of $A$. Running Algorithms \ref{Alg0} and \ref{Alg2} up to $\kappa=3$ gives the model with $\mathbf F_l=0$ for $l=2,3$ (i.e., it has a linear state-equation), with
\[\begin{array}{l}
A =  - {\rm{diag}}\left( {1.00,2.00,3.00,4.00} \right),\quad B = \left[ {\begin{array}{*{20}{c}}
1.00&0\\
0&1.00\\
1.00&0\\
0&1.00
\end{array}} \right],\\
C = \left[ {\begin{array}{*{20}{c}}
{0.50}&{ - 1.50}&{0.50}&{ - 0.50}
\end{array}} \right],\quad D = {0_{1 \times 2}},
\end{array}\]
moreover, $\mathbf H_2\in\mathbb R^{1\times 21}$, $\mathbf H_3\in\mathbb R^{1\times 56}$ are non-zero matrices, the entries of which are omitted due to space limitation.

\subsubsection{Method II}
\begin{figure*}
    \centering
    \includegraphics[width=0.8\textwidth]{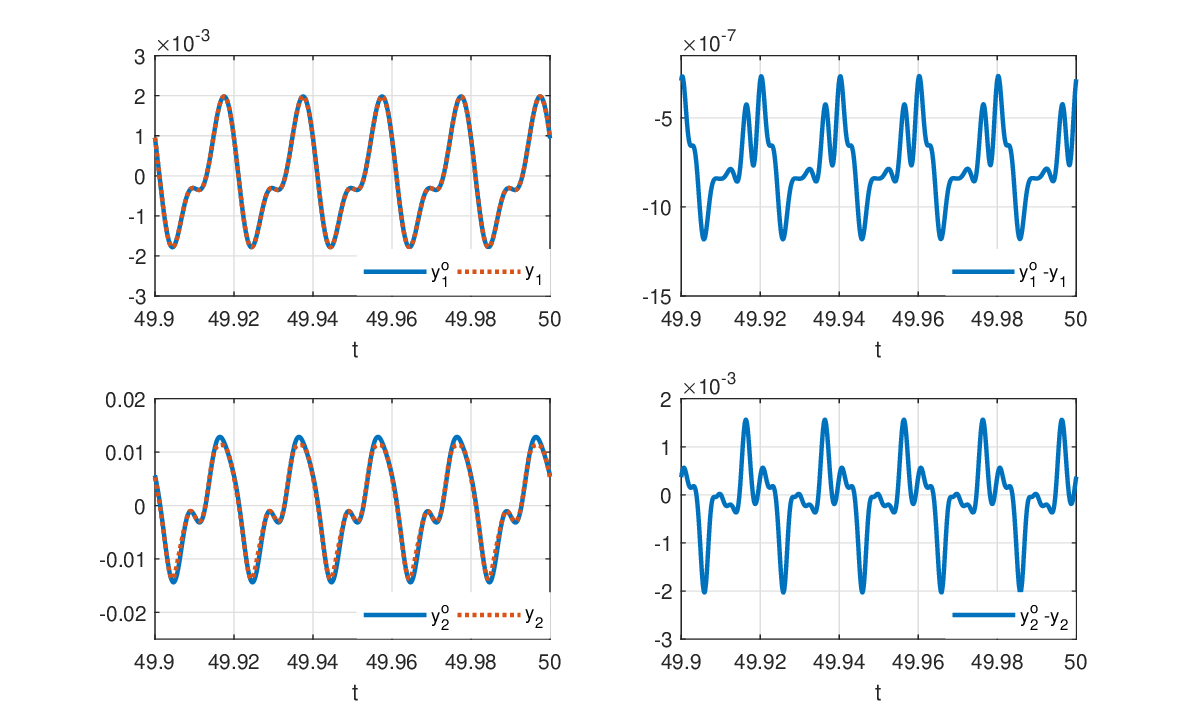}
    \caption{Method I with $D=0$: steady-state response $y^{\rm o}$ of the original model versus $y$ of the reduced model. The upper figures show the steady-state responses excited by (\ref{exo_case_II}) with initial condition $v\left(0\right)=0.2\times[1,0,1,0]^{\rm T}$ and the error between the responses; the bottom figures show the steady-state responses excited by (\ref{exo_case_II}) with initial condition $v\left(0\right)=1.2\times [1,0,1,0]^{\rm T}$ and the error between the responses.}
    \label{fig_3}
\end{figure*}

\begin{figure*}
    \centering
    \includegraphics[width=0.8\textwidth]{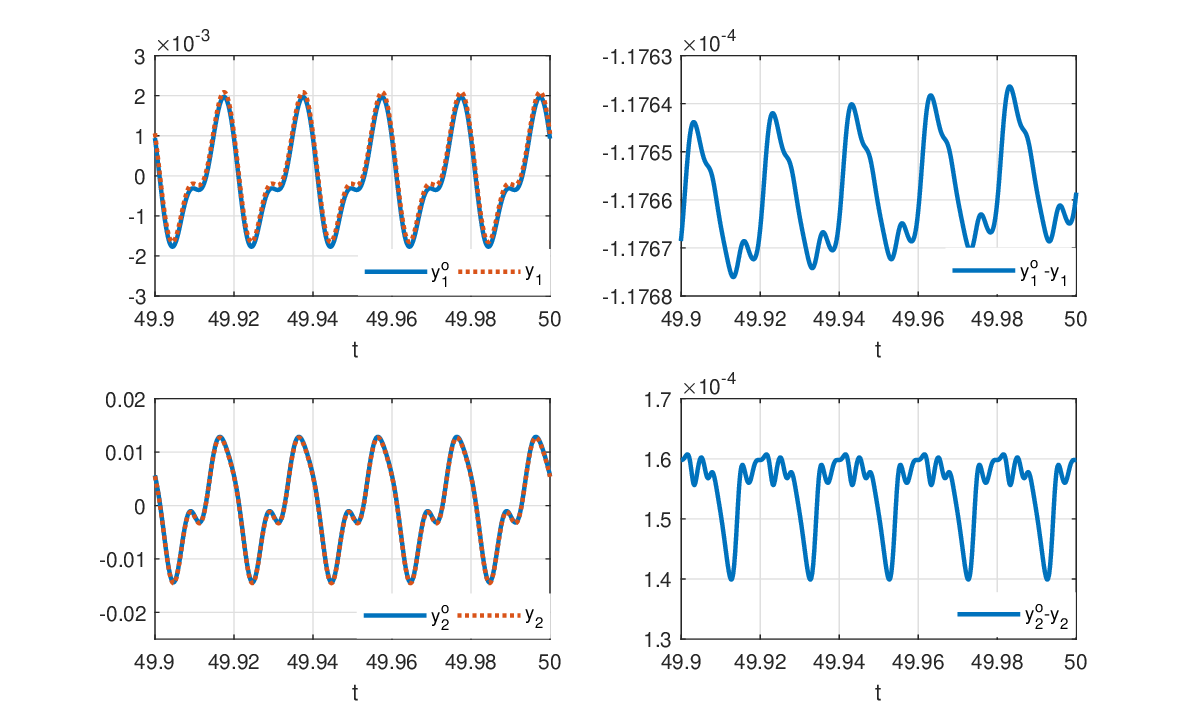}
    \caption{Method II with $D=0$: steady-state response $y^{\rm o}$ of the original model versus $y$ of the reduced model. The upper figures show the steady-state responses excited by (\ref{exo_case_II}) with initial condition $v\left(0\right)=0.2\times[1,0,1,0]^{\rm T}$ and the error between the responses; the bottom figures show the steady-state responses excited by (\ref{exo_case_II}) with initial condition $v\left(0\right)=1.2\times [1,0,1,0]^{\rm T}$ and the error between the responses.}
    \label{fig_4}
\end{figure*}

According to Theorem \ref{thm-MM_lth_general}, there exists a reduced-order model of dimension $n =2$. To obtain the model, we select $\lambda_k=-k$, $k=1,2$ as the eigenvalues of $A$. Running Algorithms \ref{Alg1} and \ref{Alg3} up to $\kappa=3$ gives
\[\begin{array}{l}
A =  - {\rm{diag}}\left( {1.00,2.00} \right),\quad B = \left[ {\begin{array}{*{20}{c}}
{0.01}&{1.00}\\
{1.00}&{ - 3.00}
\end{array}} \right],\\
C = \left[ {\begin{array}{*{20}{c}}
{1.00}&{1.00}
\end{array}} \right],\quad D = {0_{1 \times 2}},
\end{array}\]
moreover, $\mathbf H_l=0$ for $l=2,3$ (i.e., it has a linear output equation), $\mathbf F_2=[{\mathbf{F}}_{2,0},{\mathbf{F}}_{1,1},{\mathbf{F}}_{0,2}]$ where
\[\begin{array}{l}
{\bf{F}}_{2,0}={0_{2 \times 3}},\\
{{\bf{F}}_{1,1}}{\rm{ = }}\left[ {\begin{array}{*{20}{c}}
0&0&{ - 4.10 \times {{10}^{ - 2}}}&{ - 1.24 \times {{10}^{ - 3}}}\\
{5.81 \times {{10}^{ - 2}}}&0&{7.64 \times {{10}^{ - 2}}}&0
\end{array}} \right],\\
{{\bf{F}}_{0,2}}{\rm{ = }}\left[ {\begin{array}{*{20}{c}}
{ - 4.11 \times {{10}^{ - 2}}}&{1.21 \times {{10}^{ - 1}}}&{4.14 \times {{10}^{ - 2}}}\\
{4.11 \times {{10}^{ - 2}}}&{ - 1.21 \times {{10}^{ - 1}}}&{ - 4.14 \times {{10}^{ - 2}}}
\end{array}} \right],
\end{array}\]
and ${\bf{F}}_3\in\mathbb R^{2\times 20}$ is a non-zero matrix omitted for brevity. Simulation results are shown in Figures \ref{fig_3}-\ref{fig_4}. Compared to the pair in (\ref{exo_case_I}), the pair in (\ref{exo_case_II}) has relatively higher frequencies. It is observed that, in this case, the reduced-order model also performs better in fitting small signals than large signals. In this simulation, Method II shows better fitting performance for larger signals compared to Method I. Moreover, the model generated by Method II has a lower order.

\section{Conclusions and Future Works}\label{sec:Conclusions and Future Works}

The problem of moment-matching-based model reduction for high-order multi-input, multi-output polynomial nonlinear systems is addressed. A power-series decomposition technique is exploited. This enables nonlinear reduced-order models to be derived in a manner analogous to the linear case. Specifically, algorithms for determining both the order and parameters of reduced-order models with precision of degree $\kappa$ are proposed. Furthermore, new insights into the nonlinear moment-matching problem are provided: first, a lower bound for the order of the reduced-order model is obtained which, in the MIMO case, is often observed to be strictly smaller than the number of matched moments. Under mild assumptions, it is shown that a nonlinear reduced-order model can always be constructed with either a linear state equation or a linear output equation.

Future work along this direction includes considering the double-sided nonlinear moment matching problem based on power-series decomposition, and seeking a model reduction method using parameters of the original model (rather than its moments).


%

\appendix

\section{Proof of Lemma \ref{lem-equality}}\label{app_lem-equality}

Note that $n=\sigma-m$ is nothing but Eq. (\ref{Inductive_neq}) for $l=1$. Then we use the proof by induction. Assume that, for any $n,m\in\mathbb N_1$ and every $l\le k-1$ where $k\ge2$, Eq. (\ref{Inductive_neq}) holds. Then we show that Eq. (\ref{Inductive_neq}) holds for $l=k$.

For every $s,t\in\mathbb N_1$, one has \cite{Huang2024Identification}
\begin{equation}\label{Inductive_eqnC}
{\mathcal C}_{s + t}^t = {\mathcal C}_{s - 1 + t}^t + {\mathcal C}_{s - 1 + t - 1}^{t - 1} +  \cdots  + {\mathcal C}_s^1 + {\mathcal C}_{s - 1}^0,
\end{equation}
hence
\[m_l=\sum\limits_{\scriptstyle i + r = l,\hfill\atop
\scriptstyle i,r \in \mathbb N\hfill} {{\mathcal C}_{n - 1 + i}^i{\mathcal C}_{m - 1 + r}^r}  = \sum\limits_{0 \le r \le l} {{\mathcal C}_{n - 1 + l - r}^{l - r}{\mathcal C}_{m - 1 + r}^r} \]
\[ = \sum\limits_{0 \le r \le l} {\left( {{\mathcal C}_{n - 2 + l - r}^{l - r} + {\mathcal C}_{n - 2 + l - 1 - r}^{l - 1 - r} +  \cdots  + {\mathcal C}_{n - 2}^0} \right){\mathcal C}_{m - 1 + r}^r} \]
\[ = \sum\limits_{0 \le r \le l} {{\mathcal C}_{n - 2 + l - r}^{l - r}{\mathcal C}_{m - 1 + r}^r}  + \sum\limits_{0 \le r \le l - 1} {{\mathcal C}_{n - 2 + l - 1 - r}^{l - 1 - r}{\mathcal C}_{m - 1 + r}^r} \]
\[ +  \cdots  + \sum\limits_{0 \le r \le 1} {{\mathcal C}_{n - 2 + 1 - r}^{1 - r}{\mathcal C}_{m - 1 + r}^r}  + {\mathcal C}_{n - 2}^0{\mathcal C}_{m - 1}^0.\]
Then, based on the inductive assumption, one has
\[{\mathcal C}_{\sigma  - 1 + l}^l = {\mathcal C}_{\sigma  - 2 + l}^l + {\mathcal C}_{\sigma  - 2 + l - 1}^{l - 1} +  \cdots  + {\mathcal C}_{\sigma  - 1}^1 + {\mathcal C}_{\sigma  - 2}^0\]
\[ = \sum\limits_{0 \le r \le l} {{\mathcal C}_{n - 2 + l - r}^{l - r}{\mathcal C}_{m - 1 + r}^r}  + \sum\limits_{0 \le r \le l - 1} {{\mathcal C}_{n - 2 + l - 1 - r}^{l - 1 - r}{\mathcal C}_{m - 1 + r}^r} \]
\[ +  \cdots  + \sum\limits_{0 \le r \le 1} {{\mathcal C}_{n - 2 + 1 - r}^{1 - r}{\mathcal C}_{m - 1 + r}^r}  + {\mathcal C}_{n - 2}^0{\mathcal C}_{m - 1}^0=m_l.\]
This concludes the proof.

\bibliographystyle{plain}        
\bibliography{bare_jrnl}




\end{document}